\documentclass[aoas]{imsart}

\RequirePackage{amsthm,amsmath,amsfonts,amssymb}
\RequirePackage[authoryear]{natbib}
\RequirePackage[colorlinks,citecolor=blue,urlcolor=blue]{hyperref}
\RequirePackage{graphicx}

\usepackage[utf8]{inputenc} 
\usepackage[T1]{fontenc}    
\usepackage{hyperref}       
\usepackage{url}            
\usepackage{booktabs}       
\usepackage{amsfonts}       
\usepackage{nicefrac}       
\usepackage{microtype}      
\usepackage{xcolor}         

\usepackage{hyperref}
\usepackage{url}
\usepackage{bbm}
\usepackage{booktabs}
\usepackage{amsfonts}
\usepackage{amsthm}
\usepackage{bbm}

\usepackage{algorithm}
\usepackage{algpseudocode} 

\startlocaldefs
\theoremstyle{plain}

\theoremstyle{remark}

\endlocaldefs

\begin{document}

\begin{frontmatter}
\title{Quantifying uncertainty in climate projections with conformal ensembles}
\runtitle{Conformal Climate Ensembles}

\begin{aug}
\author[A]{\fnms{Trevor} \snm{Harris}\ead[label=e1,mark]{tharris@tamu.edu}},
\and
\author[B]{\fnms{Ryan} \snm{Sriver}\ead[label=e2]{rsriver@illinois.edu}}
\address[A]{Department of Statistics,
Texas A\&M University,
\printead{e1}}


\address[B]{Department of Atmospheric Sciences,
University of Illinois at Urbana Champaign,
\printead{e2}}

\end{aug}

\begin{abstract}

Ensembles of General Circulation Models (GCMs) are the primary tools for investigating climate sensitivity, projecting future climate states, and quantifying uncertainty. GCM ensembles are subject to substantial uncertainty due to model inadequacies, resolution limits, internal variability, and inter-model variability, meaning rigorous climate risk assessments and informed decision-making require reliable and accurate uncertainty quantification (UQ). We introduce conformal ensembles (CE), a new approach to climate UQ that quantifies and constrains projection uncertainty with conformal prediction sets and observational data. CE seamlessly integrates climate model ensembles and observational data across a range of scales to generate statistically rigorous, easy-to-interpret uncertainty estimates. CE can be applied to any climatic variable using any ensemble analysis method and outperforms existing inter-model variability methods in uncertainty quantification across all time horizons and most spatial locations under SSP2-4.5. CE is also computationally efficient, requires minimal assumptions, and is highly robust to the conformity measure. Experiments show that it is effective when conditioning future projections on historical reanalysis data compared with standard ensemble averaging approaches, yielding more physically consistent projections.

\end{abstract}


\end{frontmatter}

\section{Introduction} \label{sec:introduction}

\newdimen\stringwidth
\setbox0=\hbox{$\emptyset$}
\stringwidth=\wd0

General Circulation Models (GCMs) are fundamental tools for understanding the physical dynamics and interactions within the climate system, simulating historical and potential future climate change, and quantifying uncertainty \citep{kattenberg1996climate, flato2014evaluation, stocker2014climate, tebaldi2021esd, ipcc2023ch4}. Climate models are physical representations of Earth's coupled climate system that use differential equations, numerical methods, and sub-grid parameterizations \citep{randall2007climate, alizadeh2022advances}, to simulate the time evolution and interactions between the major components of the climate such as atmospheric, oceanic, land surface, sea ice, and biogeochemical based processes. GCMs can simulate plausible realizations of climate processes and variables, including their time evolution, given different initial conditions and external forcings such as changing solar activity, volcanic activity, and greenhouse gas emissions over time.  \citep{eyring2016overview}. 
Climate models are also crucial for projecting future climate states and quantifying the impacts and risks of climate change from global to regional scales \citep{ipcc2023ch10, shaw2024fc}, and under different emissions scenarios  \citep{masson2021ipcc, tebaldi2021esd, ipcc2023ch4}.


Climate modeling centers actively develop new climate models to improve their capacity to emulate the global climate system \citep{o2016scenario} and inform downstream risk assessments and mitigation efforts \citep{ipcc2023ch10}.
However, GCMs are subject to substantial internal uncertainties that do not decrease with improving model complexity \citep{deser2012uncertainty, caldwell2019doe}. These uncertainties are further compounded when producing localized projections with downscaling and bias correction techniques \citep{lafferty2023cas}. Similarly, variations between models can lead to a wide array of plausible projections, even given the same initial conditions and forcings \citep{knutti2010challenges, flato2014evaluation}. Because climate risk assessments fundamentally rely on projection uncertainty to assess the likelihood of events, robust and accurate uncertainty quantification (UQ) has become essential component of climate projection.

Current practice in climate UQ is to consider a multi-model ensemble that contains output from many different climate models run under the same time-evolving external forcings. A multi-model ensemble helps ensure that estimates of the climate response are robust across different model structures and allows us to quantify projection uncertainty through the inter-model variability (IMV) \citep{lambert2001cmip1, gleckler2008performance, knutti2010challenges, stocker2014climate}. 
The multi-model ensemble is then aggregated into a single projection, called an analysis, and paired with some notion of uncertainty, such as a 5\%-95\% model spread \citep{masson2021ipcc}. This type of analysis is ubiquitous in climate science \citep{masson2021ipcc} and is used extensively to estimate impacts and risks associated with climate change \citep{NCA2023, masson2021ipcc}.

The degree to which large ensembles represent projection uncertainty is a topic of significant recent interest \citep{jones2000managing, christensen2008need, white2013limitations, qian2016uncertainty, tett2022does}. Key sources of uncertainty in climate projections include differences in model structure, model inadequacy, and the inherent internal variability within a model \citep{hawkins2009bams, sriver2015effects, deser2020insights, tett2022does}. Uncertainties can also arise from unresolved process parameterizations, parameter value uncertainty, and even the data used for model calibration \citep{tett2022does}. These uncertainties tend to increase as climate models become more complex, resolved on finer scales, and more computationally expensive, creating challenges for projecting future regional variability and extremes \citep{shaw2024fc}. This has even led to a somewhat paradoxical situation where newer models, drawn from the Coupled Model Intercomparison Project 6 (CMIP6) \citep{eyring2016overview}, can show increased uncertainty compared to older CMIP5 models, particularly at regional scales. Therefore, because simply improving GCM complexity and resolution does not inherently reduce uncertainty \citep{raaisaanen2007reliable, deser2012uncertainty}, new approaches to uncertainty quantification (UQ) are needed to help constrain projections future projections.

Recent works have proposed various methods for reducing projection uncertainty by direct conditioning on observational data. One common approach is to use bias correction methods \citep{piani2010statistical, teutschbein2012bias, vrac2015multivariate, maraun2017towards, franccois2020multivariate} that shift and scale the ensemble to match the quantiles, or other functionals, of historical observational data. Bias correction methods are effective in reducing uncertainty in the mean states of the historical and current climate. However, they can often result in overly confident projections and are sensitive to the method and target observational data sets \citep{lafferty2023cas}. 
Similar to bias correction methods are the delta methods \citep{leeds2015simulation, poppick2016temperatures} which modify the mean, scale, or spectral density of observational data based on simulated model-based change factors between current and future climate states.

An alternative approach are Bayesian hierarchical models \citep{tebaldi2004regional, smith2009Bayesian, bhat2011climate, rougier2013second, qian2016uncertainty, sansom2017constraining, bowman2018hierarchical} that taxonomize and quantify the different sources of variability in climate models (e.g., model uncertainty, model inadequacy, and natural variability). Bayesian methods learn a posterior distribution of local or global climatologies conditional on model runs and observational data. This process yields rigorous uncertainty quantification via posterior sampling, but typically at a high computational cost, using complex models, with subjectively chosen priors. 

We propose an alternative approach to uncertainty quantification based on conformal inference \citep{vovk2005algorithmic, shafer2008tutorial}. Conformal inference is a general predictive inference methodology for imbuing any black box model with statistically valid prediction regions, even in finite samples \citep{lei2018distribution}. 
We re-cast the problem of climate projection as a prediction-like problem \citep{harris2023multimodel}, whereby an ``ensemble analysis function'', such as a deep neural network, is trained to predict observational data from a multi-model ensemble. The predictions of the analysis function constitute a synthesis of the inputted multi-model ensemble into a single projection that has been bias-corrected and downscaled against observational data. By reserving a small hold-out set of historical data, we can estimate the out-of-sample residual distribution of the analysis function and use this to form prediction sets for any future multi-model ensembles run through the same analysis function \citep{lei2018distribution}. We show that these prediction sets, represented as conformal ensembles, can be used in place of a multi-modal ensemble as a valid measure of climate projection uncertainty with significantly improved UQ skill in perfect model experiments (Section \ref{sec:data} and Section \ref{sec:application}).


Conformal ensembles offer a number of advantages over traditional large ensemble and Bayesian approaches because they do not attempt to estimate each source of uncertainty, but rather directly shortcut to the final projection uncertainty. This relatively simple scheme, however, allows us to seamlessly integrate models runs and observational data over a wide range of spatial scales, using standard statistical learning approaches, into rigorous and easy-to-understanding measures of uncertainty. Compared to traditional inter-model variability based approaches, conformal ensembles are less sensitive to the structural differences across models because their spread and statistical validity are determined by the incorporated observations. Compared to Bayesian approaches, they are computationally efficient, do not require subjective priors, and are universally applicable to any model analysis function. In fact, conformal ensembling does not require strong assumptions about the correctness of the models, the relationship between models, or the relationship between models and data. Our only requirement is that the projection residuals are exchangeable \citep{lei2014distribution} over time, or equivalently that the relationship between the models and the observations is relatively stationary over time. We demonstrate that this assumption approximately holds for many analysis functions across many climate variables (Section \ref{sec:experiments}).

\section{Ensemble analysis data} \label{sec:data}

We will use two data sources to demonstrate and apply our proposed approach. First, in Section \ref{sec:experiments}, we will use relatively small ensembles of climate models, run under historical and future forcing scenarios, to validate our approach using perfect model experiments \citep{harris2023multimodel}. This process involves jackknifing out each GCM run to serve as a surrogate observation set to validate our approach on climatologically valid simulations. This procedure will help demonstrate that a conformal inference-based approach will result in relatively high UQ skill compared to inter-model variability under a wide range of plausible "observational" scenarios. In Section \ref{sec:application}, we will then apply our approach to reanalysis data as the actual (quasi) observational process and compare the uncertainty bands generated by our approach against standard inter-model variability (IMV).

\subsection{Climate models} \label{sec:data_models}

We consider 31 member ensemble of average monthly 2-meter surface temperature (TAS) fields, a 20 member ensemble of maximum monthly 2-meter surface temperature (TMAX) fields, and a 32 member ensemble of total monthly precipitation fields (PR), all drawn from the Coupled Model Intercomparison Project (CMIP6) \citep{eyring2016overview} (Appendix). Historical climate data was run under a historical (hist.) forcing scenario, re-gridded to a $2^\circ \times 2^\circ$ grid \citep{hill2004architecture}, and truncated to January 1940 through December 2014 to match reanalysis data availability. Future climate model output was run under Shared Socioeconomic Pathway 2 and RCP 4.5 (SSP2-4.5) \citep{o2016scenario, fricko2017marker} from January 2015 through December 2099 and re-gridded to the same $2^\circ \times 2^\circ$ grid. To match reanalysis data availability (Jan 1940 - Mar. 2024), we concatenate the actual historical data (Jan. 1940 - Dec. 2014) with SSP2-4.5 output (Jan. 2015 - Mar. 2024) to form a ``historical'' train dataset for fitting ensemble analysis models and estimating conformal ensembles. We use the remaining SSP2-4.5 output (Apr. 2024 - Dec. 2099) as test data. 

The exact list of models we consider depends on the climatic variable. The superset of climate models we considered included MIROC6, MPI-ESM1-2-LR, NorESM2-LM, CMCC-ESM2, CAMS-CSM1-0, CMCC-CM2-HR4, INM-CM5-0, CIESM, IPSL-CM6A-LR, FGOALS-f3-L, INM-CM4-8, GFDL-ESM4, CAS-ESM2-0, CanESM5-1, UKESM1-0-LL, FGOALS-g3, CNRM-ESM2-1, CESM2-WACCM, KACE-1-0-G, ACCESS-ESM1-5, CMCC-CM2-SR5, NorESM2-MM, MRI-ESM2-0, NESM3, CNRM-CM6-1, FIO-ESM-2-0, KIOST-ESM, CanESM5, TaiESM1. For each climatic variable, if a given model included historical and SSP2-4.5 runs for that variable, then it was included in the ensemble. Each model is run on a 100, 250, or 500km grid under ensemble settings r1p1f1, r2p1f1, or r3p1f1.

\subsection{Reanalysis data}  \label{sec:data_reanalysis}

Instead of proper observational data, which is sparsely observed in time and space, we will use quasi-observational data, which has complete spatiotemporal coverage. Complete spatiotemporal coverage is necessary for our method to produce projection ensembles for the entire globe. Specifically, we use reanalysis data from the ERA5 reanalysis product \citep{hersbach2020era5}. Reanalysis data optimally combines observational data from a wide variety of observational sources with weather models to produce high-resolution gridded estimates. ERA5 uses Integrated Forecasting System (IFS) Cy41r2 and spans January 1940 through March 2024. We will only consider TAS, TMAX, and PR fields on single pressure levels to match the CMIP6 runs (Section \ref{sec:data_models}). However, unlike the CMIP6 runs, ERA5 interpolates on a much finer grid (30km) that does not necessarily need to downscale to match the model's resolution. A benefit of our conformal approach is that the conformal ensemble will match the resolution of the target process (reanalysis fields), which results in an automatic downscaling effect. 

We will treat the monthly aggregate ERA5 reanalysis fields as our ground truth for deriving conformal inference based uncertainty in Section \ref{sec:application}. Monthly aggregates allow us to avoid minor temporal mismatches between the reanalysis data and the models, which often do not consider leap years or are run on 360-day years. Alternative reanalysis data products, such as NCEP \citep{kalnay1996ncep}, could also be used, resulting in different uncertainty since different data products do not necessarily agree. In particular, ERA5 exhibits precipitation biases \citep{hersbach2020era5, cucchi2020wfde5} and, therefore, may not be an ideal precipitation surrogate. However, at monthly scales we consider, ERA5 does show reasonable skill consistent with alternative data products \citep{rivoire2021comparison}, particularly in extra-tropical regions \citep{lavers2022evaluation, liu2024global}. The advantage of ERA5 is its long temporal coverage (1950–2024), which provides a much longer training and calibration set than alternatives like GPCP v3.2 (1983–2023). Therefore, swapping ERA5 for NCEP or other products is not likely to strongly impact the overall results because these products are distributionally similar at monthly scales \citep{garrett2024validating}. This would impact the exact ensembles produced in Section \ref{sec:application} however.

\section{Conformal prediction sets for climate projections} \label{sec:climate_icp}

Conformal inference is a general tool for constructing exact prediction intervals for any black box algorithm. Early works focused on the univariate time series setting \citep{vovk2005algorithmic}, while later works generalized it to multivariate, functional, classification, and other settings \citep{lei2015conformal}. The most commonly used form of conformal inference is \textit{split} or inductive conformal inference, which uses sample splitting to achieve exact coverage \citep{papadopoulos2002inductive}. Algorithm \ref{alg:ICP} describes the general inductive conformal inference procedure \citep{lei2015conformal} for constructing a prediction set $C_n$ given a sequence of data $Z_1,...,Z_n$ belonging to a metric space $\Omega$, such as $\mathbb{R}^p$ for $p \geq 1$.
\begin{algorithm}
	\caption{Inductive Conformal Inference \citep{lei2015conformal}}
	\label{alg:ICP}
	\begin{algorithmic}[1]
		\State Given data $Z_1,...,Z_n \in  \Omega$, confidence level $\alpha \in (0, 1)$, and $n_1 < n$
		\State Split the data $Z_1,...,Z_{n_1}$ and $Z_{n_1 + 1},..,Z_n$. Let $n_2 = n - n_1$.
		\State Let $g : \Omega \mapsto \mathbb{R}$ be a function constructed from $Z_1,...,Z_{n_1}$
		\State Define $\sigma_i = g(Z_{n_1 + i})$ for $i \in 1,...,n_2$. Let $\sigma_{(1)} \leq ... \leq \sigma_{(n_2)}$ denote the ranked values.
		\State Return $C_n(z) = \{ z : g(z) \geq \lambda \}$ where $\lambda =  \sigma_{(\lceil(n_1+1)\alpha\rceil + 1)}$
	\end{algorithmic} 
\end{algorithm}
 Algorithm \ref{alg:ICP} is significantly more computationally efficient than the original transductive conformal method because it only requires fitting the function $g(\cdot)$ once. If we construct $C_n(z)$ according to Algorithm \ref{alg:ICP}, then given a new data point $Z_{n+1}$
\begin{equation} \label{eqn:coverage}
    1 - \alpha \leq P(Z_{n+1} \in C_n(z)) \leq 1 - \alpha + \frac{1}{n_2}
\end{equation}
for all $\alpha \in (0, 1)$ and where $n_2$ is the size of the calibration set \citep{lei2015conformal}. Equation \ref{eqn:coverage} guarantees that $C_n$ is statistically valid in finite samples and that $C_n$ is non-conservative because the upper bound $1 - \alpha + \frac{1}{n_2}$ converges to $1 - \alpha$. We will generally take $n_2 = 200$ to ensure our coverage is within $0.005$ of the nominal level. Sensitivity analysis (Section \ref{sec:sensitivity_analysis}) shows this is a reasonable and practical choice for surface temperature data, but larger calibration sets can be helpful for precipitation data.

The general split conformal inference procedure (Algorithm \ref{alg:ICP}) is fundamentally determined by $g : \Omega \mapsto \mathbb{R}$, which maps observations $z \in \Omega$ into scores $g(z) \in \mathbb{R}$.
Following \cite{lei2015conformal}, we will adapt Algorithm \ref{alg:ICP} to the climate projection setting (Algorithm \ref{alg:CCR}) by decomposing $g : \Omega \mapsto \mathbb{R}$ into a regression function $f_\theta : \mathcal{X} \mapsto \mathcal{Y}$ and a scoring rule $d : \mathcal{Y} \times \mathcal{P} \mapsto \mathbb{R}$. The regression function $f_\theta : \mathcal{X} \mapsto \mathcal{Y}$ maps climate model ensembles $x \in \mathcal{X}$ to observational fields $y \in \mathcal{Y}$, while the scoring rule $d : \mathcal{Y} \times \mathcal{P} \mapsto \mathbb{R}$ maps predictions $\hat y  \in \mathcal{Y}$ and their empirical distribution $\mathcal{P}$ to real-valued scores.
We will denote this class of regression functions as ``multi-model ensemble analysis'' functions because they combine multi-model ensembles into a single prediction. We will use a specific class of scoring rules called depth functions.

\subsection{Multi-model ensemble analysis} \label{sec:multi_model_analysis}

Multi-model ensemble analysis is the practice of combining a collection of climate model projections into a single, unified projection, called an ensemble analysis.
The most straightforward and common approach to ensemble analysis is to take the pointwise ensemble mean as the combined projection. However, ensemble averaging typically shows sub-optimal projection skill \citep{flato2014evaluation} compared to methods that condition on observational data \citep{flato2014evaluation, vrac2015multivariate, abramowitz2019esd, harris2023multimodel}. Regardless of the ensemble analysis method, unconditional or conditional, we will treat the ensemble analysis as the output of a regression function trained to predict observational data given an ensemble of climate model runs \citep{vrac2015multivariate, harris2023multimodel}.

To specify an appropriate class of regression functions for multi-model ensemble analysis, we introduce the following notation to formalize our setting. Let $X_{t, i} \in \mathbb{R}^{q_i}$, $q_i \geq 1$, denote the output of climate model $i \in 1,...,m$ at time $t \in 1,...,n$, and let $Y_t \in \mathbb{R}^{p}$, $p \geq 1$, denote observations from the true climate process at time $t$. Each $X_{t, i} \in \mathbb{R}^{q_i}$ is a gridded field, possibly having different resolutions $q_i = (q^{lat}_i \times q^{lon}_i)$, where $q^{lat}_i$ and $q^{lon}_i$ denote the number of latitude and longitude points on the grid.  Each $Y_t \in \mathbb{R}^{p}$ is also a gridded field with resolution $p = (p^{lat} \times p^{lon})$. We use $X_t = \{X_{t, 1},...,X_{t, m}\} \in \mathbb{R}^q$ to denote the output of an ensemble of $m$ climate model runs at time $t$, where $q = \sum_{i = 1}^m q_i$. We use $Z_t = (X_t, Y_t)$ to refer to the climate model ensembles and the corresponding observational field observed at time $t$.


Algorithm \ref{alg:ICP} allows us to construct sets $C_\alpha(X) \subset \mathbb{R}^p$ such that 
\begin{equation}
    P(Y_t \in C_{\alpha}(X_t)) \geq 1 - \alpha,
\end{equation}
where $P$ is the joint distribution of $Z_t = (Y_t, X_t)$ \citep{lei2018distribution}. To do this, we require an ensemble analysis function $f_\theta$ that can map the ensemble to the observations.
We define a multi-model ensemble analysis function as any function 
\begin{equation}\label{eqn:analysis_function}
    f_\theta: \mathbb{R}^q \mapsto \mathbb{R}^p,
\end{equation}
possibly parameterized by $\theta \in \Theta$, that maps the ensemble $X_t \in \mathbb{R}^q$ to the observations $Y_t \in \mathbb{R}^p$. This definition includes commonly used analysis methods, such as the ensemble mean and weighted ensemble means \citep{giorgi2002calculation, giorgi2003probability, flato2014evaluation, abramowitz2019esd}. It also allows for more complex regression models such as local linear regressions, gaussian process regression, and deep neural networks \citep{harris2023multimodel}. 

Equation \ref{eqn:analysis_function} allows for enormous flexibility in defining the multi-model ensemble analysis function because, essentially, any parametric or nonparametric model is permitted. Given this flexibility, we should prefer regression models $f_\theta: \mathbb{R}^q \mapsto \mathbb{R}^p$ that are as accurate as possible. The more precise $f_\theta$ is, the sharper the conformal prediction sets can be without sacrificing coverage \citep{lei2018distribution}. Also, because $f_\theta: \mathbb{R}^q \mapsto \mathbb{R}^p$ has to learn a high-dimensional regression function on limited training data (typically $q,p \gg n$), we should prefer models that are robust to the curse of dimensionality \citep{jacot2024dnns}.

If our chosen model $f_\theta$ contains learnable parameters, i.e. $\theta \neq \emptyset$, then we can train $f_\theta$ as in any other regression problem. Let $f_\theta \in \mathcal{F}_\Theta$ denote a class of models, let $D_n = \{(X_t, Y_t)\}_{t = 1}^n$ denote a training set of $n$ historical realizations of $X_t$ and $Y_t$, and let $\mathcal{L}:(\theta, D_n) \mapsto \mathbb{R}$ denote an appropriate loss function over the parameters $\theta \in \Theta$. We estimate the optimal $\hat f_\theta$ as
$\hat f_\theta = \arg\min_{\theta \in \Theta} \mathcal{L}(\theta, D_n)$
and make predictions as $\hat Y = \hat f_\theta(X)$ for any $X \in \mathbb{R}^q$.


\subsection{Scoring multi-model ensemble forecasts}
\label{sec:multi_model_conformity}
Regardless of the chosen ensemble analysis function, the trained model $\hat f_\theta$ returns a multivariate prediction that must be scored and ranked according to Algorithm \ref{alg:ICP}. Typically, scoring and ranking are performed by measuring the magnitude of the residuals of the predictions, i.e., $||y_t - \hat f_\theta(X_t)||$ for some norm $||\cdot||$, and ordering these values from least to greatest. Here, we propose a more general approach based on the classical concept of data depth, a nonparametric notion for imposing a center out ordering of the observations \citep{liu1990notion, zuo2000general, mosler2013depth, nagy2016integrated}. We will treat the residual process, $R_t = Y_t - \hat f_\theta(X_t)$, as a random process, $R_t \sim \mathcal{R}$ and use a specific depth function to order $R_1,R_2,...$ from closest to the centroid of $\mathcal{R}$ to furthest from the centroid.

A depth function implements a specific notion of depth, for example, the halfspace depth, simplicial depth, projection depth, zonoid depth, $\ell_2$-Depth, and $\Phi$-Depths among others \citep{mosler2013depth}. Let $y \in \mathbb{R}^p$ be a vector and $P \in \mathcal{P}$ a distribution, a valid depth function is any function 
\begin{equation}\label{eqn:depth}
d:  \mathbb{R}^p \times \mathcal{P} \mapsto [0, 1],
\end{equation}
mapping $(y, P) \mapsto d(y \mid P)$, that satisfies the depth postulates (D1-D5) \citep{mosler2013depth}. The depth postulates essentially require $d:  \mathbb{R}^p \times \mathcal{P} \mapsto [0, 1]$ to be translationally invariant, scale equivariant, and decrease monotonically from the centroid of the distribution. 
Any depth functions will satisfy the depth postulates to a varying degree and can be used to rank observations from most central (highest depth value) to least central (lowest depth value). The differences between depth functions are mainly in their computational tractability and robustness due to the different ways they characterize outlyingness.
In Section \ref{sec:coverage}, we show how two different depth functions lead to more or less conservative empirical size control for the corresponding conformal prediction set.

We use depth functions for scoring and ranking because they more accurately represent the level sets of $\mathcal{R}$ compared to simple norms. A norm, or more accurately its inverse, can act like a type of depth that assumes $\mathcal{R}$ is centered at $0 \in \mathbb{R}^p$ with ball-shaped (depending on the norm) level sets. Proper depth functions allow for biased residuals and convex, or in some cases star-shaped, level sets \citep{mosler2013depth, nagy2016integrated}. As noted in \cite{lei2018distribution}, a better approximation of the level sets of $\mathcal{R}$ results in more accurate prediction sets, a fact we empirically observe in Table \ref{tab:coverage}.

Any depth function $d(\cdot)$ can approximate the level sets of $\mathcal{R}$ with its depth defined central regions. Because depths decrease monotonically from the centroid, any number $\tau \in [0, 1]$ defines a convex central region 
$$
D_\tau = \{r \in \mathbb{R}^p : d(r, \mathcal{P}) \geq \tau \}
$$
that includes all points, $r \in \mathbb{R}^p$ with a depth value greater than $\tau$. Central regions are key to how we implement the general ICP algorithm (Algorithm \ref{alg:ICP}) for climate model projections (Algorithm \ref{alg:CCR}). We will, essentially, just take $\tau$ to be the $\lceil(n_1+1)\alpha\rceil + 1)$ smallest depth value and use the $D_\tau$ central region, translated by $\hat f_\theta$, as our prediction region $C_n(Z)$ (Algorithm \ref{alg:ICP} step 5). This procedure is the ordinary conformal prediction procedure for regression \citep{lei2018distribution}, but for high-dimensional multivariate targets using data depth for ranking.

\subsection{Conformal prediction sets for climate fields} \label{sec:conf_ensembles}



Given a sufficiently adequate multi-model ensemble analysis function $f_\theta \in \mathcal{F}_\Theta$, where $f_\theta: \mathbb{R}^q \mapsto \mathbb{R}^p$, and a valid depth function $d:  \mathbb{R}^p \times \mathcal{P} \mapsto [0, 1]$ (Equation  \ref{eqn:depth}), we can implement the general ICP algorithm (Algorithm \ref{alg:ICP}) for climate projection.
Using the notation from Section \ref{sec:multi_model_analysis}, suppose we have a historical dataset $D_{\text{hist}} = \{ (X_i, Y_i) \}_{i = 1}^n$, an analysis function $f_\theta$, a depth function $d$, a confidence level $\alpha \in (0, 1)$, and an out of sample input ensemble $X_{n+1}$. Algorithm \ref{alg:CCR} describes our procedure for computing $C_{\alpha}(X_{n+1})$ such that $P(Y_{n+1} \in C_{\alpha}(X_{n+1})) \geq 1-\alpha$.
\begin{algorithm}
	\caption{Conformal Central Region}
	\label{alg:CCR}
	\begin{algorithmic}[1]
		\State Partition the historical data $D_{\text{hist}}$ into disjoint training and calibration sets $$ D_{\text{train}} = \{(X_t, Y_t)\}_{t = 1}^{n_1} \quad D_{\text{cal}} = \{(X_t, Y_t)\}_{t = n_1 + 1}^n,$$ and let $n_2 = n - n_1$ denote the size of the calibration set.
		
		\State Train the model $f_\theta(\cdot)$  with loss $\mathcal{L}$ on $D_{\text{train}}$ as $$\hat \theta = \arg\min_{\theta \in \Theta} \mathcal{L}(f_\theta, D_{\text{train}}).$$
		
		\State Compute $R_t = Y_t - f_{\hat \theta}(X_t)$ on $D_{\text{cal}}$ and let $\mathcal{R}_{\text{cal}}$ denote the distribution of $R_{n_1 + 1},..., R_{n}$.
		
		\State Compute the depths of the residual fields $R_t$ with respect to $\mathcal{R}_{\text{cal}}$
$$\mathcal{D}_{cal} = d(R_{n_1 + 1} \mid \mathcal{R}_{\text{cal}}),..., d(R_n \mid \mathcal{R}_{\text{cal}}).$$

		\State Compute $\tau = Quantile(\mathcal{D}_{cal}, \lceil  (n_2+1)(1-\alpha)  \rceil)/n_2)$ to generate the residual central region $$D_{\alpha}(R) = \{R \sim \mathcal{R}_{\text{cal}} : d(R \mid \mathcal{R}_{\text{cal}}) \geq \tau \}$$
		
		\State Return the prediction central region for $Y_{n+1}$, $$C_{\alpha}(X_{n+1}) = \{f_{\hat \theta}(X_{n+1}) + R : R \in D_{\alpha}(R) \}$$ 
	\end{algorithmic} 
\end{algorithm}

Lines 1, 2, and 3 of Algorithm \ref{alg:CCR} follow the standard split conformal inference procedure. We split the full historical dataset into two disjoint subsets $D_{\text{train}}$ and $D_{\text{cal}}$, train our multi-model ensemble analysis function $f_\theta(\cdot)$ on $ D_{\text{train}}$ and compute the residuals on $D_{\text{cal}}$. The calibration set $D_{\text{cal}}$ is necessary for properly measuring the distribution of the residuals since the residuals on the train set $ D_{\text{train}}$ will be overly shrunk towards zero due to overfitting. Lines 4 and 5 use our chosen depth function $d(\cdot)$ to find the $\tau = (n_2+1)(1-\alpha)  \rceil)/n_2)$ central region of the out-of-sample residuals. The level $\tau$ is, essentially, a finite sample correction of the nominal coverage $1 - \alpha$ to ensure exact empirical coverage. Finally, in line 6, we return the prediction region for $Y_{n+1}$ by simply translating the residual central region by the prediction $f_{\hat \theta}(X_{n+1})$.

Algorithm \ref{alg:CCR} can be interpreted as ``converting'' the multi-model ensemble of climate model runs observed at time $n+1$, $X_{n+1}$, into an exact conformal prediction set $C_{\alpha}(X_{n+1})$. We take an inexact approximation of the projection distribution, model ensembles, and convert it into a constrained approximation, conformal ensembles. This approach is similar to Conformalized Quantile Regression (CQR) methods \citep{romano2019conformalized, kivaranovic2019adaptive} that convert arbitrary prediction quantiles into exact conformal quantiles. However, because the depth function $d(\cdot)$ measures outlyingness over the entire spatial domain simultaneously, the resulting projection sets $C_{\alpha}(X_{n+1})$ are jointly valid over $Y_{n+1}$ instead of only pointwise valid.

\subsection{Conformal ensembles for uncertainty quantification} \label{sec:conf_projection}
The standard approach to representing multivariate or functional prediction sets $C_{\alpha}(X_{n+1})$, generated by Algorithm \ref{alg:CCR}, is to compute the pointwise minimum and maximum bounds that form a prediction band \citep{lei2015conformal, diquigiovanni2021importance}. Bands are commonly used because they are computationally efficient and form a convex hull around the level sets of $\mathcal{R}_{cal}$. However, because bands are a loose approximation, they contain many projections that do not belong to $C_{\alpha}(X_{n+1})$, and they discard information about the distribution of the ensemble. This makes bands less than ideal for climate projection, where we do not want our prediction sets to contain highly implausible climate fields and we want to assess the projection distribution.



We propose an alternative characterization of $C_{\alpha}(X_t)$ based on ensembles rather than bands. That is, instead of focusing on the boundaries of $C_{\alpha}(X_t)$, we will focus on the density of the prediction inside $C_{\alpha}(X_t)$. We can do this by sampling an ensemble of forecasts $\hat Y_t^1,...,\hat Y_t^m \in C_{\alpha}(X_t)$ to generate an ensemble forecast for $Y_t$. This will allow us to assess how well the projections of different ensemble analysis functions concentrate around the target process using traditional UQ metrics such as the Continuous Ranked Probability Skill score (CRPS) \cite{gneiting2007strictly} or distributional distance such as the Wasserstein distance \citep{bonneel2015sliced,villani2021topics}. Ensemble forecasts are ubiquitous throughout climatological forecasting because they allow for an indirect characterization of complex forecast distributions. By taking a conformal approach to ensemble construction, we can ensure that the ensemble is jointly calibrated over the entire spatial domain.

We convert $C_{\alpha}(X_t)$ into an ensemble forecast by reusing the $\lceil (n_2+1)(1-\alpha) \rceil$ least outlying residuals, $R_{(1)},...,R_{\lceil (n_2+1)(1-\alpha) \rceil)}$ that generated $C_{\alpha}(X_t)$ as a kind of deterministic sample of $C_{\alpha}(X_t)$. Given these ``generating'' residuals we can construct a $1-\alpha$ conformal ensemble $E_\alpha(X)$ as 
\begin{equation} \label{eqn:conformal_ensemble}
E_\alpha(X_t) = \{\hat f_\theta(X_t) + R_{(i)} \}_{i = 1}^{\lceil (n_2+1)(1-\alpha) \rceil},
\end{equation}
where $R_{(i)}$ is the $i$th smallest residual in $\mathcal{R}_{cal}$ according to the depth function $d(\cdot)$. The conformal ensemble $E_\alpha(X)$ is fully contained inside $C_\alpha(X)$, so our ensemble forecast will not contain any projections outside the theoretical projection set, unlike a band might. Furthermore, given our ensemble analysis function and depth function, this ensemble is trivial to compute. Once we determine the $\lceil (n_2+1)(1-\alpha)  \rceil$ least outlying residuals (Line 5 Algorithm \ref{alg:CCR}) we can simply add them to any future projection $\hat f_\theta(X_{n+1})$ to generate an ensemble forecast. 

Algorithm \ref{alg:CCR} and our ensemble representation provide a simple and convenient means for constructing statistically valid projection ensembles. We can treat $E_\alpha(X)$ as a drop-in replacement for the model ensemble $X_t$ that is well calibrated, significantly larger (provided $n_2 \gg m$), and shrinks the more accurate we make $f_\theta$. However, this does not guarantee that $E_\alpha(X)$ will not contain climatologically implausible realizations. In the following section (Section \ref{sec:experiments}), we assess the validity of the projection sets and the UQ skill of the projection ensembles.

\section{Experiments}  \label{sec:experiments}

We conduct ``perfect model'' experiments to assess the coverage and UQ skill of the proposed conformal approach across a range of climate variables, ensemble analysis functions, and depth functions \citep{knutti2017climate, harris2023multimodel}. In each experiment we have an ensemble of $m$ climate model runs (TAS: $m = 31$, TMAX: $m = 20$, PR: $m = 30$), which we split into a train set (Jan. 1940 - Jun. 2007), calibration set (Jul. 2007 - Mar. 2024), a test set (Apr. 2024 - Dec. 2099). Iterating over all $m$ model runs, we treat one run as the target process, or stand-in, for observational data and use the remaining $m - 1$ model runs as our multi-model ensemble. We train six different ensemble analysis functions and construct, for each, three different conformal ensembles based on three different depth functions on train and calibration sets respectively (Section \ref{sec:data_models}). We compare the UQ skill of each combination against the inter-model variability (IMV) and bias-corrected inter-model variability (IMV (BC)) \citep{vrac2015multivariate} on the test set. By repeating this procedure for all $m$ model runs, we can assess conformal uncertainty quantification across a wide range of plausible climate trajectories.

The six ensemble analysis functions we consider include the pointwise average (EA), a weighted pointwise average (WA) \citep{bishop2013climate}, the delta method (Delta), a pointwise linear model (LM) \citep{abramowitz2015climate}, a Gaussian process regression using a neural-network Gaussian process (NNGP) kernel (GP) \citep{harris2023multimodel}, and a deep convolutional neural network (CNN). WA and LM are the same models, except in WA, the weights are trained jointly, and there is no intercept term, whereas, in LM, the weights are learned independently at each spatial location. We used an NNGP kernel for the GP because it was shown to be more robust to covariate shift than standard isotropic kernels \citep{harris2023multimodel}. For the CNN, we used five layers with a width of 64 and 3x3 convolutions with a skip connection between the 2nd and 4th layers. The CNN was trained with the Adam optimizer \citep{kingma2014adam} using a constant learning rate of $10^{-3}$ and no regularization, which was found to have no measurable impact on the results.

The three depth functions we consider include the $\ell_{\infty}$-Depth based on the $\ell_\infty$-norm, the Integrated Tukey depth (Tukey), and the (inverse) $\ell_\infty$ norm (norm), which is a non-depth based scoring rule. The $\ell_{\infty}$-Depth of a vector $r$ with respect to distribution $P$ is defined as $D_\infty(r, P) = 1/(1 + \mathbb{E}[||r - R||_\infty])$ where $R \sim P$. The Integrated Tukey depth averages the univariate Tukey depth $D_{\text{Tukey}}(r, P)  = \mathbb{E}[\min\{ \hat F_x(r(x), 1 - \hat F_x(r(x)) \}]$ over each element of $r$. Here $\hat F_x$ denotes the empirical CDF of the residuals at location $x$. Finally, the $\ell_\infty$-norm simply computes $D(x) =1/(1 + ||x||_\infty)$. We choose $\ell_\infty$-norm based depths and scoring rules, instead of the usual $\ell_1$ or $\ell_2$ norm based measures because $\ell_\infty$ was found to be more robust to changes in the distribution of the ensembles. In all cases, the resulting conformal ensemble could be generated in under a second on either an NVIDIA V100 or an NVIDIA 4700 GPU.

\subsection{Coverage and sharpness} \label{sec:coverage}

We first assess the ability of the conformal method to empirically achieve the nominal coverage under our six ensemble analysis functions, three depth measures, three climatic variables, and a pure white noise ``control run''. We measure the performance of each combination by its empirical coverage compared to the nominal level $\alpha = 0.1$, and the average width of the prediction set. The width of the prediction is approximated as the average distance between the upper and lower bands (Section \ref{sec:conf_projection}) of the conformal ensemble. Each entry in Table \ref{tab:coverage} shows the average empirical coverage and width over all perfect model experiments within each variable.

The white noise control run does not use any of the climate model data (Section \ref{sec:data_models}), but instead simulates pure white noise fields from a standard Gaussian distribution. We sample ``ensembles'' of size $m = 30$ on an $30 \times 50$ grid, with 800 train observations, 200 validation observations, and 1000 test observations. We report the coverage on the test set averaged over the $m = 30$ experiments. The white noise control runs are to verify that our approach works, regardless of the ensemble analysis and depth function, when exchangeability holds.
After verifying on white noise, we then assess the empirical coverage on CMIP6 model data (TAS, TMAX, PR). TAS is the monthly average surface temperature, TMAX is the monthly maximum surface temperature, and PR is the (log) monthly total precipitation. TAS and PR are high-priority climate variables used extensively in climate change studies (IPCC), while TMAX allows us to assess our approach to extremal processes. Unlike the white noise runs, each of these variables exhibit distribution shift due to climate change (SSP2-4.5 forcings  \citep{o2016scenario}).

\begin{table}[!t]
\centering
\resizebox{\columnwidth}{!}{
\begin{tabular}{l|cccc|cccc}
\toprule
\midrule
Metric & \multicolumn{4}{l}{Empirical Coverage ($\alpha = 0.1$)} & \multicolumn{4}{l}{Average Width ($\downarrow$)} \\
\midrule
Variable & WN & TAS & TMAX & PR  & WN & TAS & TMAX & PR \\
\midrule
EA ($\ell_\infty$) & 0.902 & 0.890 & 0.893 & 0.906 & 6.126 & 3.717 & 3.532 & 2.611 \\
EA (Tukey) & 0.900 & 0.928 & 0.929 & 0.929 & 6.100 & 3.717 & 3.534 & 2.617 \\
EA (Norm) & 0.901 & 0.870 & 0.831 & 0.906 & 6.120 & 3.705 & 3.524 & 2.612 \\
\midrule
WA ($\ell_\infty$) & 0.900 & 0.921 & 0.922 & 0.910 & 6.220 & 3.563 & 3.474 & 2.884 \\
WA (Tukey) & 0.900 & 0.967 & 0.964  & 0.938 & 6.195 & 3.563 & 3.472 & 2.887 \\
WA (Norm) & 0.901 & 0.933 & 0.925 & 0.920 & 6.213 & 3.562 & 3.474 & 2.885\\
\midrule
Delta ($\ell_\infty$) & 0.902 & 0.890 & 0.893 & 0.906 & 6.126 & 3.731 & 3.605 & 2.871 \\
Delta (Tukey) & 0.900 & 0.928 & 0.929 & 0.929 & 6.100 & 3.731 & 3.605 & 2.877 \\
Delta (Norm) & 0.900 & 0.904 & 0.893 & 0.906 & 6.120 & 3.730 & 3.604 & 2.872 \\
\midrule
LM ($\ell_\infty$) & 0.902 & 0.900 & 0.913 & 0.912 & 6.323 & 5.341 & 4.831 & 3.121 \\
LM (Tukey) & 0.903 & 0.975 & 0.969 & 0.945 & 6.330 & 5.346 & 4.837 & 3.128 \\
LM (Norm) & 0.901 & 0.874 & 0.868 & 0.903 & 6.332 & 5.328 & 4.816 & 3.119 \\
\midrule
GP ($\ell_\infty$) & 0.902 & 0.932 & 0.933 & 0.908 & 6.126 & 3.376 & 3.295 & 2.805 \\
GP (Tukey) & 0.900 & 0.977 & 0.975 & 0.949 & 6.100 & 3.375 & 3.293 & 2.809 \\
GP (Norm) & 0.900 & 0.931 & 0.926 & 0.916 & 6.120 & 3.374 & 3.295 & 2.806 \\
\midrule
CNN ($\ell_\infty$)  & 0.900 & 0.876 & 0.856 & 0.901 & 6.353 & 3.686 & 3.646 & 2.952 \\
CNN (Tukey) & 0.905 & 0.936 & 0.935 & 0.928 & 6.326 & 3.687 & 3.644 & 2.958 \\
CNN (Norm) & 0.894 & 0.895 & 0.882 & 0.909 & 6.347 & 3.684 & 3.645 & 2.955 \\
\midrule
\bottomrule
\end{tabular}
}
\caption{Out of sample (Apr. 2024 - Dec 2099) empirical coverage, and average width, of each model analysis function (EA, WEA, Delta, LM, GP,  CNN) on TAS, TMAX, and PR and a white noise (WN) control run averaged over all perfect model experiments. We included a white noise control run to assess the empirical coverage of conformal inference when all assumptions are met. Climate variable results show that conformal prediction controls the out-of-sample coverage near the nominal level, with a minor few exceptions.  Proper depth functions ($\ell_{\infty}$-Depth and Tukey depth) show nominal to slightly conservative coverage.}
\label{tab:coverage}
\end{table}

Table \ref{tab:coverage} shows that our method achieves (nearly) perfect coverage on the white noise setting (WN) across all analysis functions (EA, WA, Delta, LM, GP, CNN) and depth functions ($\ell_{\infty}$, Tukey, Norm). WN experiments demonstrate that when historical and future data are drawn from identical distributions, the conformal method will result in statistically exact prediction sets regardless of the choice of depth function. Furthermore, the resulting ensembles' average width is largely the same ($\approx 6.1)$ in each case, indicating that very similar ensembles each method selecs a similar ensemble.

The empirical coverage for each combination applied to TAS, TMAX, and PR (Table \ref{tab:coverage}) deviates from the nominal level due distribution shift as expected. However, we can see that these violations are not too severe (lowest: TAS: 0.870, TMAX: 0.831, PR: does not undercover) and, in general, result in slight to moderate over coverage (highest: TAS: 0.977, TMAX: 0.975, PR: 0.949). However, these violations suggest that some depth functions may be more robust to distribution shift than others, and some may be overly conservative. The most severe under-coverage violations are under the Norm depth function and improper depth function, while the most severe over-coverage violations are under the Tukey depth. In fact, the Tukey depth consistently over covers the target across all variables. In terms of coverage, the most consistently precise depth function is the $\ell_{\infty}$-Depth.

However, the average widths (Table \ref{tab:coverage}) suggest that over and under coverage may be fairly benign for this problem. The resulting conformal ensembles all have roughly the same width for each depth function within a given variable and ensemble analysis function, except LM. For example, EA (Tukey) on TMAX (Cover: 0.929, Width: 3.534) and EA (Norm) on TMAX (Cover: 0.831, Width: 3.524), over and under cover the target, respectively, but have an average width difference of only 0.01. Thus, they are largely generating the same conformal ensemble, the object we use to measure UQ, even if the theoretical prediction sets are, respectively, more or less conservative.

The results in Table \ref{tab:coverage} highlight the importance of choosing an accurate ensemble analysis function. In the WN experiment, all methods could equally learn the underlying regression function, thus resulting in nearly identical residual distributions with similar widths. However, there can be significant variation in the widths of different ensemble analysis functions for the climatic variables, which are assumed to have more complex underlying regression functions. Comparing EA (average width: 3.71) against the GP (average width: 3.37), we see that the more accurate GP method resulted in narrower prediction sets. This is a natural consequence of the conformal approach being based on the residuals of out-of-sample predictions.

\begin{figure}[!t]
\begin{center}
\includegraphics[width=0.95\textwidth]{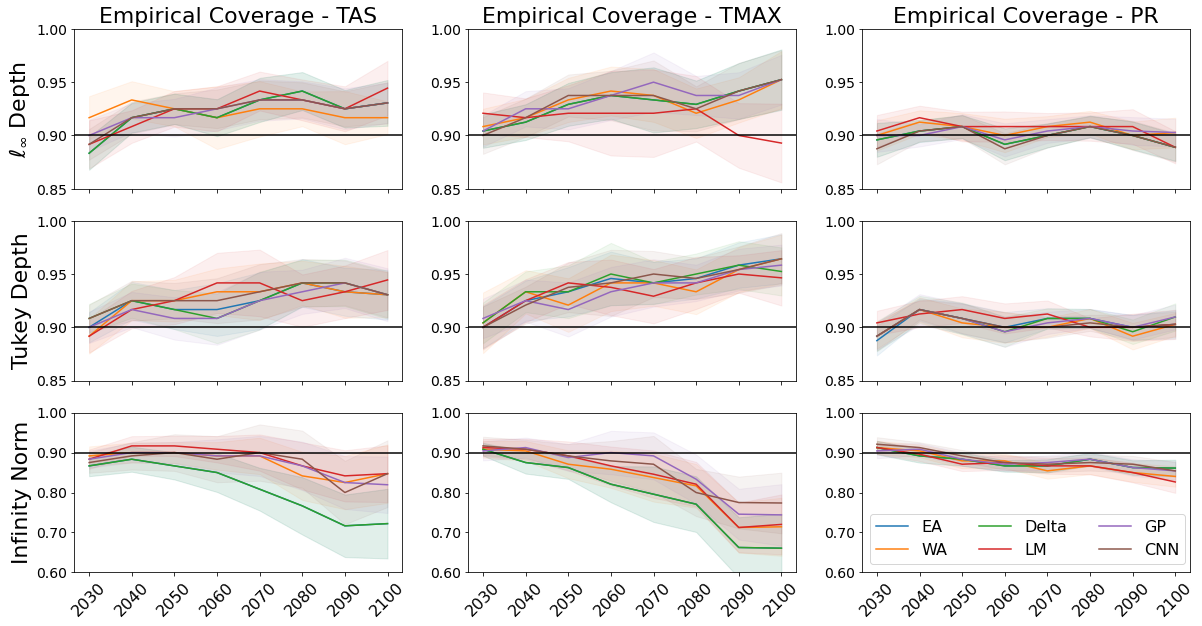} 
\caption{Empirical coverage of the ($\alpha$ = 0.1) conformal prediction set generated by each depth function, for each ensemble analysis function and climatic variable. Solid lines show the average empirical coverage for a given analysis function across all perfect model experiments and shading represents $\pm$2 standard errors. Proper depth functions ($\ell_{\infty}$-Depth and Tukey depth) show consistent nominal to slightly conservative coverage, while the improper depth function (Infinity Norm) undercovers in the latter part of the prediction horizon (2070-2100).}
\label{fig:temporal_coverage}
\end{center}
\end{figure}

Figure \ref{fig:temporal_coverage} decomposes the results in Table \ref{tab:coverage} to show the coverage of each combination over time. We compute the coverage for each combination within (approximately) decadal blocks (2024-2030, 2030-2040,...,2090-2100) to illustrate how coverage is impacted by distribution shift in each variable. The $\ell_{\infty}$-Depth is highly stable around, or above, $0.9$ across all ensemble analysis functions and variable types. This suggests that is slightly conservative and thus relatively unaffected by distribution shift within each variable. As was seen in Table \ref{tab:coverage}, the Tukey depth tends to be overly conservative and becomes more conservative over time in TAS and TMAX. The Infinity Norm is not robust against distribution shift, as seen by the dramatic loss of coverage in TAS and TMAX.

\subsection{Uncertainty Quantification} \label{sec:uq_skill}

Using our approach, we can construct a conformal prediction set and a conformal ensemble (Section \ref{sec:conf_projection}). While the prediction set is theoretical, the conformal ensemble can be used in practice for uncertainty quantification (UQ). We compare the UQ skill of our approach, using each ensemble analysis and depth function combination, against the standard ensemble inter-model variability (IMV) and a quantile corrected version of the inter-model variability (IMV(BC)) \citep{vrac2015multivariate}. The IMV and IMV(BC) are re-centered in each experiment from the ensemble mean to the ensemble analysis function's projection.

Our primary evaluation metric is the Sliced Wasserstein distance (SW) \citep{bonneel2015sliced}. SW is a distance function defined between probability distributions on a metric space defined as
$$
SW(P, Q) = \left( \int_{\mathbb{R}^{n-1}} W_2(\mathcal{R}_\theta[P], \mathcal{R}_\theta [Q]) d_\theta \right)^{1/2}
$$
where $P, Q$ are distributions on an $n$-dimensional metric space, $\mathcal{R}_\theta[P]$ is a one dimensional projection of $P$, and $W_2$ is the ordinary Wasserstein distance based on the $\ell_2$-norm \citep{villani2021topics}. The one-dimensional Wasserstein distance is
$$
W_2(F, G) = || F^{-1} - G^{-1}||_2,
$$
i.e. the $\ell_2$-norm between the quantiles of two 1D distributions, $F$ and $G$. Using SW, we can quantify the distance between the joint distribution of the projection (ensemble analysis with uncertainty arising from either a conformal ensemble or a multi-model ensemble) and the observed target process. SW will be zero if and only if projection and the target process follow the same distribution, and it will monotonically increase the further they are apart. SW is a powerful but underutilized tool for high-dimensional forecast evaluation because it compares the entire joint distribution of the projection with the target. In our experiments, we will frequently denote this distance as the ``climatology distance'' because it compares the projected climatology against the target climatology over the given time period.

\begin{table}[!t]
\centering
\resizebox{\columnwidth}{!}{
\begin{tabular}{l|cccccc}
\toprule
TAS & EA & WA & Delta & LM & GP & CNN  \\
\midrule
IMV & 1.245 (0.054) & 1.277 (0.055) & 1.249 (0.055) & 1.292 (0.055) & 1.288 (0.056) & 1.281 (0.057)  \\
IMV (BC) & 1.007 (0.067) & 1.011 (0.067) & 1.008 (0.067) & 1.042 (0.067) & 1.023 (0.067) & 1.014 (0.068)  \\
CE ($\ell_\infty$) & 0.805 (0.043) & 0.758 (0.036) & 0.793 (0.040) & 0.900 (0.033) & 0.749 (0.034) & 0.739 (0.031)  \\
CE (Tukey) & 0.796 (0.041) & 0.751 (0.034) & 0.784 (0.038) & 0.893 (0.031) & 0.742 (0.032) & 0.730 (0.029)  \\
CE (Norm) & 0.801 (0.042) & 0.755 (0.037) & 0.789 (0.039) & 0.899 (0.033) & 0.746 (0.034) & 0.737 (0.031)  \\
\midrule
TMAX & EA & WA & Delta & LM & GP & CNN  \\
\midrule
IMV & 1.048 (0.054) & 1.066 (0.053) & 1.051 (0.054) & 1.062 (0.052) & 1.068 (0.052) & 1.064 (0.053)  \\
IMV (BC) & 0.826 (0.054) & 0.837 (0.053) & 0.827 (0.054) & 0.856 (0.051) & 0.841 (0.052) & 0.837 (0.053)  \\
CE ($\ell_\infty$) & 0.676 (0.041) & 0.692 (0.040) & 0.675 (0.041) & 0.777 (0.035) & 0.678 (0.039) & 0.678 (0.041)  \\
CE (Tukey) & 0.673 (0.039) & 0.692 (0.037) & 0.672 (0.039) & 0.778 (0.033) & 0.678 (0.038) & 0.676 (0.039)  \\
CE (Norm) & 0.675 (0.041) & 0.690 (0.039) & 0.673 (0.041) & 0.775 (0.035) & 0.676 (0.040) & 0.675 (0.041)  \\
\midrule
PR & EA & WA & Delta & LM & GP & CNN  \\
\midrule
IMV & 0.222 (0.012) & 0.228 (0.012) & 0.222 (0.012) & 0.230 (0.012) & 0.222 (0.012) & 0.225 (0.012)  \\
IMV (BC) & 0.182 (0.006) & 0.186 (0.006) & 0.182 (0.006) & 0.192 (0.006) & 0.184 (0.006) & 0.182 (0.006)  \\
CE ($\ell_\infty$) & 0.157 (0.004) & 0.161 (0.004) & 0.156 (0.004) & 0.168 (0.004) & 0.157 (0.004) & 0.157 (0.004)  \\
CE (Tukey) & 0.158 (0.004) & 0.162 (0.004) & 0.158 (0.004) & 0.170 (0.004) & 0.158 (0.004) & 0.158 (0.004)  \\
CE (Norm) & 0.155 (0.003) & 0.160 (0.004) & 0.155 (0.004) & 0.168 (0.004) & 0.155 (0.004) & 0.156 (0.004)  \\
\midrule
\bottomrule
\end{tabular}
}
\caption{Sliced Wasserstein distance between the projected distribution and the target process for each model analysis function and UQ method over the projection horizon (Apr. 2024 - Dec 2099). Each number represents the average SW distance across all perfect model experiments and the number in parenthesis is one standard error. Regardless of the analysis function, depth function, or climatic variable, conformal ensembles (CE) show lower projection error compared to the IMV and IMV(BC).}
\label{tab:uq_metrics}
\end{table}

Table \ref{tab:uq_metrics} shows the average SW distance of each combination (analysis function + depth function, or analysis function + IMV or IMV(BC)), plus one standard error computed over all $m$ perfect model experiments. Regardless of the ensemble analysis function (EA, WEA, Delta, LM, GP, CNN), variable type (TAS, TMAX, PR), or depth function, swapping out the inter-model variability (IMV) for a conformal ensemble (CE) results in lower average SW distance to the target process. For example, on average, switching from IMV to CE ($\ell_{\infty}$) results in a 37.8\% decrease in SW distance in TAS, a 34.8\% decrease in TMAX, and a 29.1\% decrease in PR. The same holds even if the IMV is bias-corrected using quantile correction, although the reduction is less. These results imply that CE provides a near uniform improvement in quantifying uncertainty compared to the IMV and the IMV(BC), since there are no combinations where any CE ensemble is outperformed by the IMV or IMV (BC).

A major concern with bias correcting methods and other alternative approaches to UQ is the problem of overconfidence. We pose that the improvements seen in Table \ref{tab:uq_metrics}, using true depth functions ($\ell_{\infty}$-Depth and Tukey), are not likely caused by overconfidence of the CE ensemble because, as was shown in Table \ref{tab:coverage}, conformal methods have well controlled empirical coverage. The CE ensemble is drawn uniformly from the $1-\alpha$ confidence region (Section \ref{sec:conf_ensembles}), so it is not simply concentrated near the median. In fact, the results in Table \ref{tab:coverage} suggest that the conformal prediction sets may even be slightly underconfident because the empirical sizes are generally greater than the nominal level. CE (Norm) shows slightly improved metrics compared with $\ell_{\infty}$-Depth and Tukey depth, but due to its consistent under-coverage towards the latter part of the prediction interval (2070-2100), this may be due to overconfidence.

\subsection{Uncertainty quantification over time} \label{sec:temporal_stability}

As in Section \ref{sec:coverage}, we break the disaggregate UQ skill results in Table \ref{tab:uq_metrics} into decadal blocks to assess UQ skill over time. Because conformal methods can be susceptible to distribution shift, this will allow us to identify if there are any periods where CE performs worse than IMV or IMV(BC). We then further decompose our temporal results into monthly units, where SW is computed using only prediction sets and observations within a given month to assess seasonal variability. For CE, we use the $\ell_{\infty}$-Depth due to its overall more stable coverage (Figure \ref{fig:temporal_coverage}) and comparable UQ skill compared to the other depths. 

\begin{figure}[!t]
\begin{center}
\includegraphics[width=0.95\textwidth]{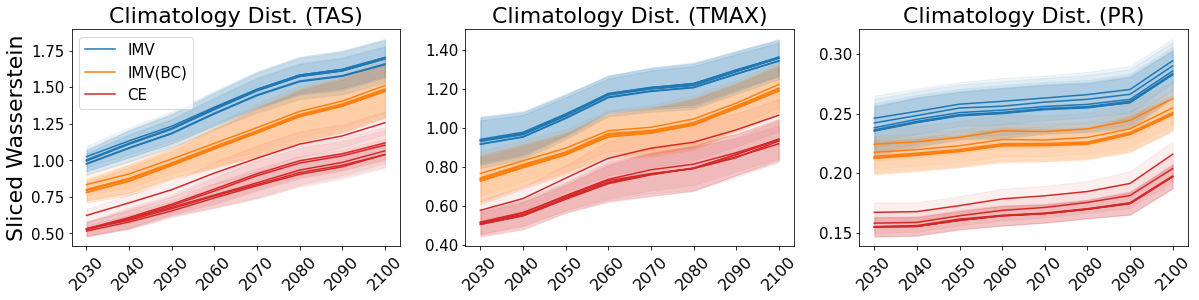} 
\includegraphics[width=0.95\textwidth]{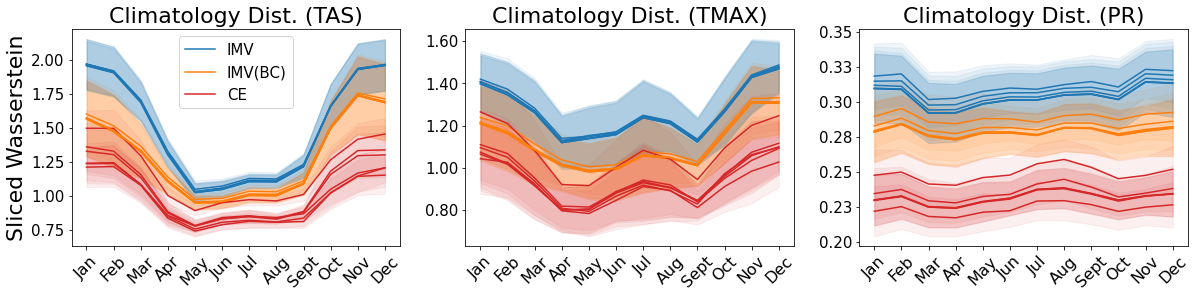} 
\caption{\textbf{Top:} Sliced Wasserstein (SW) distance between the projected distribution and the target distribution for each analysis function and UQ method within each decade of the projection period (May 2024 - Dec 2099). Each solid line represents a different analysis function and shading around each line represents $\pm$2 standard errors. \textbf{Bottom:} Same as the top row except SW was computed within each month, instead of each decade, over the projection period. For both plots we compute CE using the $\ell_{\infty}$-Depth.}
\label{fig:temporal_stability}
\end{center}
\end{figure}

The top row of Figure \ref{fig:temporal_stability} shows the SW distance between the projection and the target over a given decade using each UQ method (IMV, IMV(BC), and CE) for each ensemble analysis function averaged over all model experiments. Although SW metrics tend to increase over time, particularly for TAS and TMAX projections, CE is consistently below IMV and IMV(BC) for all models and all time periods. Thus, even for far future projections (2070 - 2100), the climatologies produced by CE are much closer to the climatologies of the target process than if IMV had been used. One notable exception is LM using CE, which sits higher than all other models using CE in TAS and TMAX. This was likely caused by instabilities training LM, resulting in overly high or low projections at a small handful of spatial locations. 

The bottom row of Figure \ref{fig:temporal_stability} instead looks at the average SW distance between the projection and the target within a given month to assess seasonal variability. This decomposition of the SW distance reveals that, while the results are much closer than in the decadal breakdown, CE still results in consistently lower SW compared to IMV and IMV(BC). Thus, the monthly climatologies of the projection based on CE tend to be closer to the observed monthly climatologies than if IMV or IMV(BC) had been used. 

Figure \ref{fig:temporal_stability} also shows that CE is sensitive to distribution shift, as indicated by the steadily increasing sliced Wasserstein distance over time. However, because the rate of increase is the same across all UQ methods and analysis functions, this suggest that the source of this sensitivity is likely in the underlying analysis function. If the analysis function is sensitive to distribution shift then it will exhibit large errors which translates to poor distribution approximation, regardless of the chosen UQ method. I.e. a statistical model using CE based UQ does not show greater sensitivity to distribution shift than the baseline approach: an ensemble average using IMV based UQ.

\subsection{Marginal uncertainty quantification over space} \label{sec:marginal_uq}

Table \ref{tab:uq_metrics} shows that, regardless of the ensemble analysis function or climate variable, swapping out IMV or IMV(BC) for a conformal ensemble (CE) uniformly decreases the approximation error of the projected climatology to the target processes climatology. Figure \ref{fig:temporal_stability} confirms that this result holds for each decade in the projection interval (Apr. 2024 - Dec 2099). Because these results are based on the Sliced Wasserstein (SW) distance, they measure how close the projection's joint distribution is to the target process' joint distribution. They do not tell us how close their distributions are marginally at each grid location, i.e., if there are any spatially varied differences. 

\begin{figure}[!t]
\begin{center}
\includegraphics[width=0.95\textwidth]{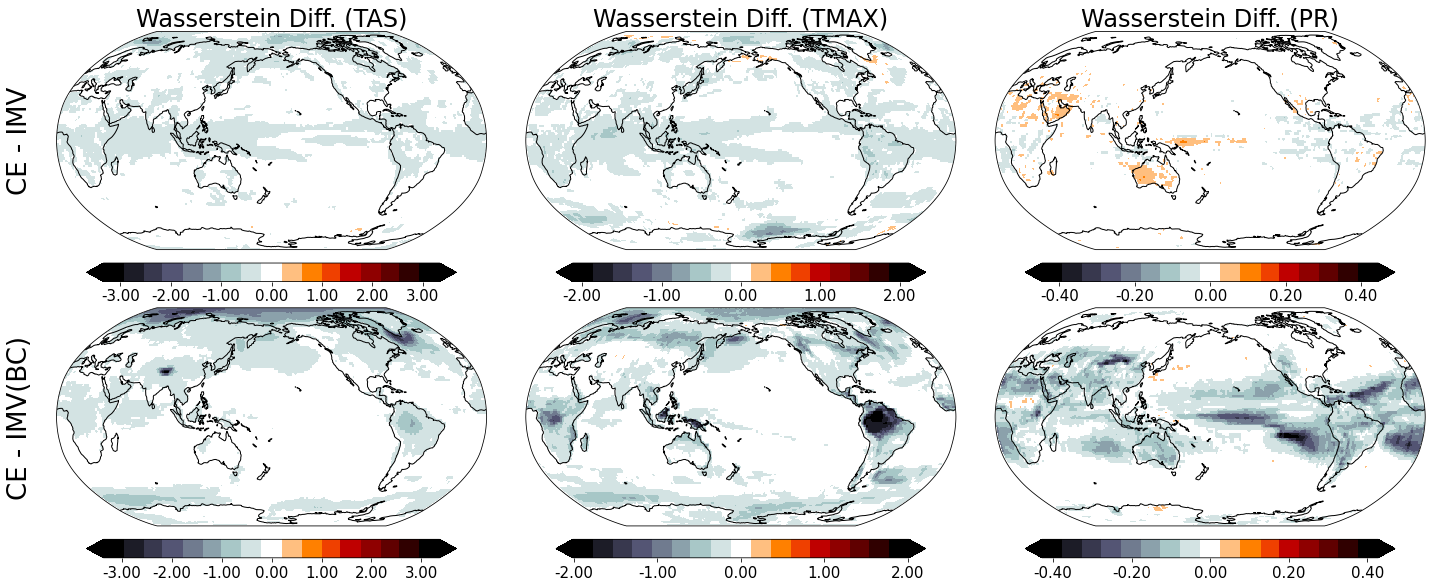} 
\caption{\textbf{Top:} Average pointwise Wasserstein distance (to the target process) across all ensemble analysis functions and model experiments using CE (with $\ell_{\infty}$-Depth) minus the same using IMV to quantify uncertainty. Grey areas indicate where the CE better approximates the target distribution compared to IMV or IMV(BC). \textbf{Bottom:} Same as the top, except using IMV(BC) to quantify uncertainty. CE shows consistent improvement at high latitudes for TAS and TMAX compared to IMV and IMV(BC).}
\label{fig:spatial_stability}
\end{center}
\end{figure}

To measure marginal differences between the projected distribution and the target process distribution, i.e., between their marginal climatologies at each location, we apply the 1D Wasserstein distance, based on the based on the $\ell_2$-norm, at each location over the entire prediction interval. Then, to compare the relative performance of CE against IMV and IMV(BC), we created the difference maps (Wasserstein Diff.) maps in Figure \ref{fig:spatial_stability}, which subtract the Wasserstein distance achieved by a GP using the IMV (or IMV(BC)) from the Wasserstein distance achieved by a GP using the CE based on the $\ell_{\infty}$-Depth. These results are then averaged over all perfect model experiments. We denote the analysis function + UQ combinations as GP + IMV, GP + IMV(BC), and GP + CE, respectively. We chose the GP as our baseline model due to its strong performance in Table \ref{tab:uq_metrics}.

The top row of Figure \ref{fig:spatial_stability} shows the Wasserstein difference maps for GP+ CE and GP + IMV. Grey areas represent spatial locations where GP + CE has a lower Wasserstein distance than GP + IMV, indicating GP + CE better approximates the marginal climatology. Red areas indicate that GP + IMV better approximates the marginal climatology, and white areas indicate little to no difference. For TAS, the biggest improvements are seen at tropical latitudes and in the Arctic, whereas the mid-latitudes (north and south) show little to no difference. TMAX shows a similar pattern except with a large improvement around the Amundsen Sea near Antarctica. There is little to no difference for PR, except in Western Australia, the Arabian peninsula, and Micronesia, where the IMV provides a small but consistent improvement. These three regions suggest that the IMV may handle low and high precipitation extremes better than the CE at these locations, despite its low resolution.

The bottom row of Figure \ref{fig:spatial_stability} shows the Wasserstein difference maps for GP+ CE and GP + IMV(BC). Compared to  GP + IMV(BC), for TAS, there are fewer improvements across the tropical latitudes but bigger improvements near the Arctic, Tibet, the Amazon rainforest, and the Southern Ocean. TMAX shows a similar pattern as TAS, except for even larger improvements, especially around the Amazon rainforest and the Southern Ocean. For PR, GP + CE strongly improves across the tropical and mid-latitudes, particularly around the Intertropical convergence zone (ITCZ). Compared with GP + IMV, GP + IMV(BC) has significantly worse performance in both extremely dry regions (Sahara desert and Arabian peninsula) and extremely wet regions (ITCZ). These results suggest that while IMV(BC) improves the joint distribution (Table \ref{tab:uq_metrics}) compared to IMV, it sacrifices its marginal approximation capabilities, particularly for extremes (TMAX and high and low regions of PR). CE, however, improves over the joint distribution of the IMV and generally results in equal or superior marginal skill. 

\subsection{Sensitivity Analysis} \label{sec:sensitivity_analysis}

We conducted a sensitivity analysis to investigate how sensitive the conformal method is to the size of the calibration set. Intuitively, a larger calibration set should allow for a more complete exploration of the predictive distribution. However, a larger calibration set leaves less data for training the integration function because we have a fixed total sample size. This can deteriorate the out-of-sample predictive skill, resulting in overly wide prediction sets. We evaluate the EA and GP methods on increasingly larger calibration sets, again using their Sliced Wasserstein (SW) distance on the test set.

\begin{figure}[!t]
\begin{center}
\includegraphics[width=0.99\textwidth]{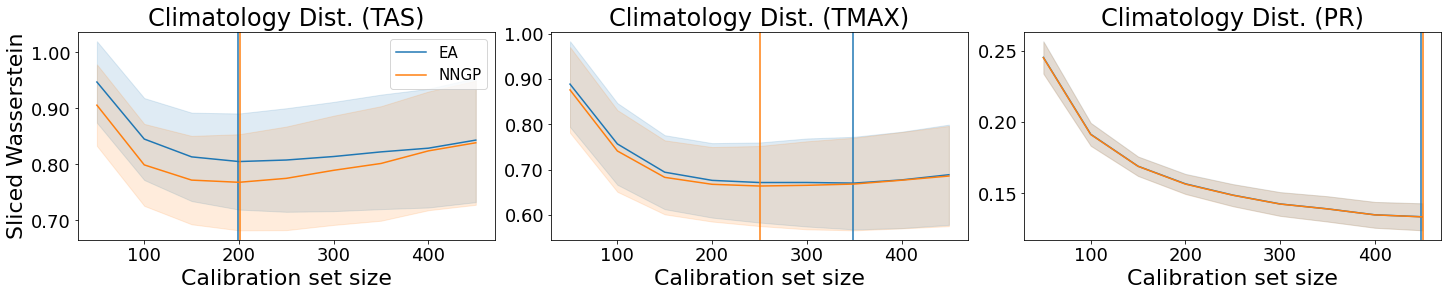} 
\caption{Blue lines show the mean, over all perfect model experiments, sliced Wasserstein distance between EA + CE and the target model over the test period (May 2024 - Dec 2099) using an increasingly large calibration set size. The orange line shows the same except using GP + CE. Blue and orange shading represents $\pm$2 standard errors from the mean.}
\label{fig:calibration_sensitivity}
\end{center}
\end{figure}

Figure \ref{fig:calibration_sensitivity} shows the out-of-sample test SW distances for EA and GP across 10 different calibration size settings ($n_2$ = 50, 100, ..., 450) and each climatic variable TAS, TMAX, and PR. Each line represents the median SW distance across all perfect model experiments, and the uncertainty bands represent 2 standard errors (2SE) from the median. As expected, UQ skill shows a ``U'' shaped performance curve where small and large calibration sets result in poor test UQ. For the GP method, this result is natural, considering a larger calibration set leaves less data for training. 
We included the EA method to show that this effect is not entirely due to insufficient training data. Because the EA method has no training step, its similarly shaped performance curve indicates that larger calibration sets might include residual functions that are irrelevant for future prediction. Again, this result is somewhat expected because the distributions of TAS, TMAX, and PR evolve over time. By including residuals too far in the past, our projection set might contain information that is irrelevant or even counter to future projections. Thus, even if sample size was not a factor, for non-stationary climatic variables, we may still need to limit the size or time period of the calibration set.

\section{Application} \label{sec:application}

Projecting climate variability is essential for quantifying the risks and impacts of global climate change under a wide range of emissions and socioeconomic scenarios. Ordinarily, large multi-model ensembles, such as those in the Coupled Model Intercomparison Project (CMIP6) \citep{eyring2016overview}, are used to quantify projection uncertainty through their inter-model variability (IMV). Our approach, based on conformal inference, allows us to condition these ensembles on quasi-observational data, such as reanalysis data products (Section \ref{sec:data}), to improve their theoretical and empirical uncertainty quantification (UQ) skill (Section \ref{sec:experiments}) without being overconfident (Table \ref{tab:coverage}). We will use our conformal method (Section \ref{sec:climate_icp}) to construct projection ensembles and compare them against an ensemble of climate models, with and without bias correction. Again, we consider the two most widely studied climatic variables: temperature and precipitation fields on single pressure levels.

Specifically, we use each of the ensemble analysis functions from Section \ref{sec:experiments} (EA, WA, Delta, LM, GP, CNN) to project future monthly average surface temperatures (TAS) and monthly total precipitation (PR) under Shared Socioeconomic Pathway 2 and RCP 4.5 (SSP2-4.5) \citep{o2016scenario}. We again use the CMIP6 climate model ensembles (described in Section \ref{sec:data_models}) as input to each ensemble analysis function, but now we will use them to predict reanalysis fields (described in Section \ref{sec:data_reanalysis}). As in section \ref{sec:experiments}, we train each ensemble analysis function on historical data and estimate the conformal ensemble on an out-of-sample calibration set. However, because our sensitivity analysis (Figure \ref{fig:calibration_sensitivity}) showed that PR strongly benefited from a larger calibration set, we used a calibration set of size $n_2 = 200$ for TAS and size $n_2 = 400$ for PR. Thus, the analysis function will be trained on Jan. 1940 - Jun. 2007 historical data for TAS and on Jan. 1940 - Nov. 1990 historical data for PR. The calibration sets will, therefore, cover the periods Jul. 2007 - Mar. 2024 for TAS and  Dec 1990 - Mar. 2024 for PR. 

\subsection{Historical validation} \label{sec:historical_validation}

We first compare IMV and IMV(BC) against CE (using $\ell_{\infty}$-Depth) on a small hold-out set to verify that the advantages of CE seen in the simulation experiments (Table \ref{tab:uq_metrics} still hold on real data. For TAS, we train each ensemble analysis function on historical data (Jan. 1940 - June 2007), estimate the conformal ensemble on the first half of the calibration set (Jul. 2007 - Nov. 2015), then compute the sliced Wasserstein distance between projections and the reanalysis data in the second half of the calibration set (Dec. 2015 - Mar. 2024). For PR, we will do the same except we train from Jan. 1940 - Nov. 1990, estimate the ensemble from Dec. 1990 - June 2007 and then check the sliced Wasserstein metric from Jul. 2007 - Mar. 2024. Essentially, for this experiment only, we will mimic the experiments in Table \ref{tab:uq_metrics}, using reanalysis data by splitting the calibration sets in half and using the first half as our new calibration set and the second half as our test set.

\begin{figure}[!t]
\begin{center}
\includegraphics[width=0.975\textwidth]{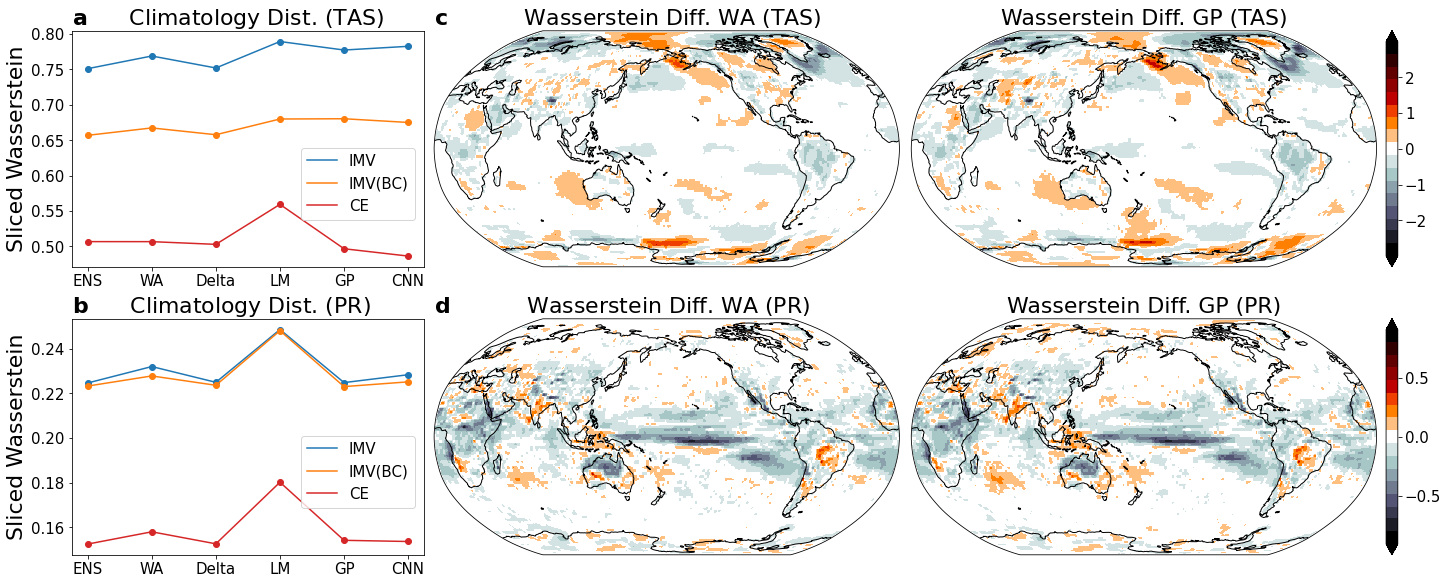} 
\caption{Panels (\textbf{a}) and (\textbf{b}) show the sliced Wasserstein distance between the projected distribution, for each model plus UQ combination, and the reanalysis data on the held out period for TAS and PR, respectively. Panel (\textbf{c}) shows the pointwise Wasserstein distance (to the observations) of WA + CE($\ell_{\infty}$) minus the pointwise Wasserstein distance (to the observations) of EA + IMV on the left, and the same using GP + CE($\ell_{\infty}$) on the right. Panel (\textbf{d}) is the same as panel (\textbf{c}) except computed on PR fields. Grey areas indicate where our method better approximates the observational distribution.}
\label{fig:reanalysis_validation}
\end{center}
\end{figure}

The top row of figure \ref{fig:reanalysis_validation} shows the out-of-sample sliced Wasserstein distance between the projections (with uncertainty) and reanalysis fields on the held out data for TAS. As in Section \ref{sec:experiments}, we consider all combinations of ensemble analysis functions and UQ methods and, again, see that CE results in uniformly lower sliced Wasserstein distances than either IMV or IMV(BC). To compare the marginal projected distributions across grid points, we include two Wasserstein Diff. maps, as in Section \ref{sec:marginal_uq}, that compare the compare the marginal distribution of WA + CE against EA + IMV and GP + CE against EA + IMV. We use EA + IMV as the baseline because it is the default choice for UQ in climate projections. Areas in red indicate that EA + IMV more closely matches the reanalysis distribution, while areas in grey indicate WA + CE or GP + CE more closely matches the reanalysis distribution. Remarkably, these are nearly the same plot, implying that the deficiencies in the marginal distributions WA + CE and GP + CE may not be due to model choice (WA vs GP) but due to our limited ability to estimate the CE with only 100 observations. 

The second row of \ref{fig:reanalysis_validation} shows the out-of-sample sliced Wasserstein distance between the projections (with uncertainty) and reanalysis fields on the held out data for PR. The first plot shows that using CE results uniformly lower SW numbers than either IMV or IMV(BC) across all ensemble analysis functions. Again, the Wasserstein Diff. maps are remarkably similar to each other and show that CE has a much better marginal approximation of the PR distribution in the Tropical Pacific near the Intertropical Convergence Zone (ITCZ). This region tends to feature variability on multiple time scales annual cycle (ITCZ), interseasonal variability (e.g., ENSO), and long-term trends due to global warming, all of which global climate models can struggle to emulate.

\subsection{Forced response in the tails} \label{sec:spatial_extremes}

Given the positive results in Section \ref{sec:historical_validation}, confirming that CE provides reliable UQ for reanalysis data, we use our method to project TAS and PR fields over the next 30 years (Apr. 2024 - Dec. 2054). We use the GP model as our ensemble analysis function, trained on Jan. 1940 - Nov. 2015 data for TAS and Jan 1940 - Nov. 1990 data for PR, and construct CEs on Jul. 2007 - Mar. 2024 data for TAS and Dec 1990 - Mar. 2024 data for PR. Unlike in Section \ref{sec:historical_validation}, we can use the full calibration set to construct each CE. Using the GP, we make three projections over all months between 2024 and 2054 using the IMV, IMV(BC), and the CE to quantify uncertainty and denote these combinations as GP + IMV, GP + IMV(BC), and GP + CE, respectively. 

Because each combination, GP + IMV, GP + IMV(BC), and GP + CE, produces the same mean prediction, we primarily investigate how each UQ method affects the upper (95\%) and lower (5\%) quantiles of the projected distribution. Specifically, we compute the 95\% and 5\% quantiles of each projection at each grid location over the 30 year projection period and compare these with the 95\% and 5\% quantiles of the historical reanalysis data at each grid location over the 30 year period Jan 1960 - Dec. 1989. Our goal is to show how each projection method forecasts changes in the distribution's tails, from a given reference period, to show how each UQ method projects changes in the tails due to climate change.

\begin{figure}[!t]
\begin{center}
\includegraphics[width=0.975\textwidth]{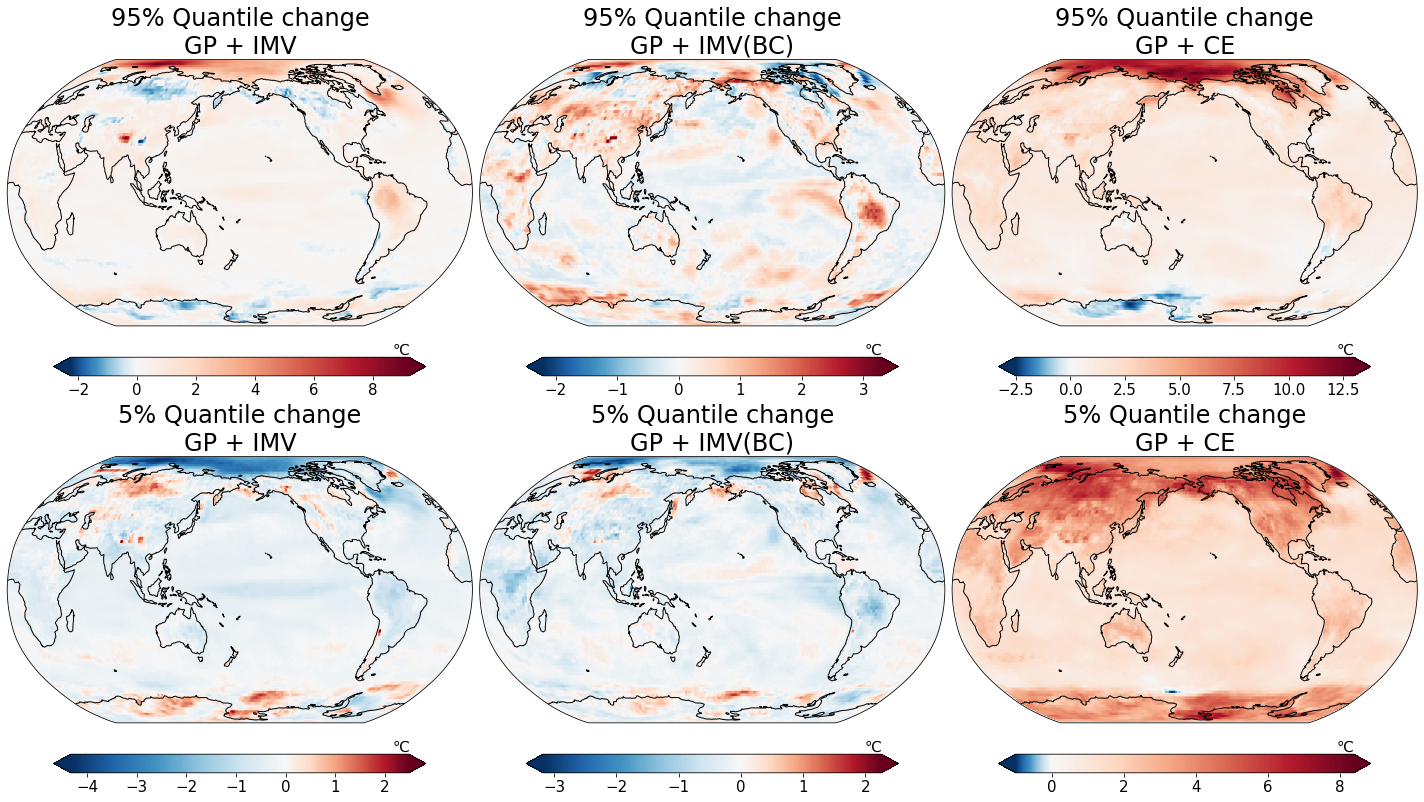} 
\caption{\textbf{Top:} Pointwise 95\% quantile difference maps for a Gaussian process model with each of the three UQ methods on monthly average surface temperature (TAS) data. Maps show the difference between the projected 95\% quantile for TAS over the period 2024-2054 and the historical 95\% quantile for TAS over the period 1960-1990 at that location. \textbf{Bottom:}  Same as the top row but using 5\% quantiles instead. Red areas indicate the 95\% or 5\% quantile has increased relative to the baseline historical period, and blue areas indicate a decrease.}
\label{fig:tas_extremes}
\end{center}
\end{figure}

The top row of Figure \ref{fig:tas_extremes} shows the 95\% quantile change plots for TAS using the three projection methods. Each location on these plots shows the 95\%  quantile of the 30-year projection minus the 95\% quantile of the 30-year historical period. Red areas indicate that the 95\% quantile has increased over the reference period, while blue areas indicate that the 95\% quantile has decreased. Both GP + IMV and GP + CE show heavy warming in the Arctic, around $6 - 8^\circ$C and $10 -12^\circ$C, respectively, and slight cooling in the Southern Ocean near Antarctica, around $0-2^\circ$C. Both methods also show slight warming in the rest of the ocean. Overall, the changes in the 95\% quantile between GP + IMV and GP + CE are broadly consistent with each other, with GP + CE tending to show a more consistent increase with particularly extreme increases in the Arctic. GP + IMV(BC) shows similar spatial patterns of warming and cooling as GP + IMV but with a highly restricted range, projecting at most a $3^\circ$ warming. This effect is likely due to quantile mapping overfitting the tails of the historical data \citep{berg2024robust, rahimi2024future}.

The bottom row of Figure \ref{fig:tas_extremes} shows the 5\% quantile change plots for TAS, in which GP + CE and GP + IMV heavily disagree. GP + CE shows consistent increases in the 5\% quantile, indicating that it anticipates the lower range of the temperature distribution to increase dramatically over the next 30 years, particularly in high and low latitudes. This result is predicted from atmospheric physics based on reductions in the meridional temperature gradient, which climate models often struggle to capture \citep{ipcc2023ch4}. Together with the 95\% quantile plot, this indicates that GP + CE is projecting an overall shift upwards in the temperature distribution. Conversely, GP + IMV is less certain about the projected distribution, with the 5\% quantile maps showing sustained decreases, particularly in the Arctic ($2-4^\circ$C). Again, GP + IMV(BC) is spatially consistent with GP + IMV but with a reduced range.

\begin{figure}[!t]
\begin{center}
\includegraphics[width=0.975\textwidth]{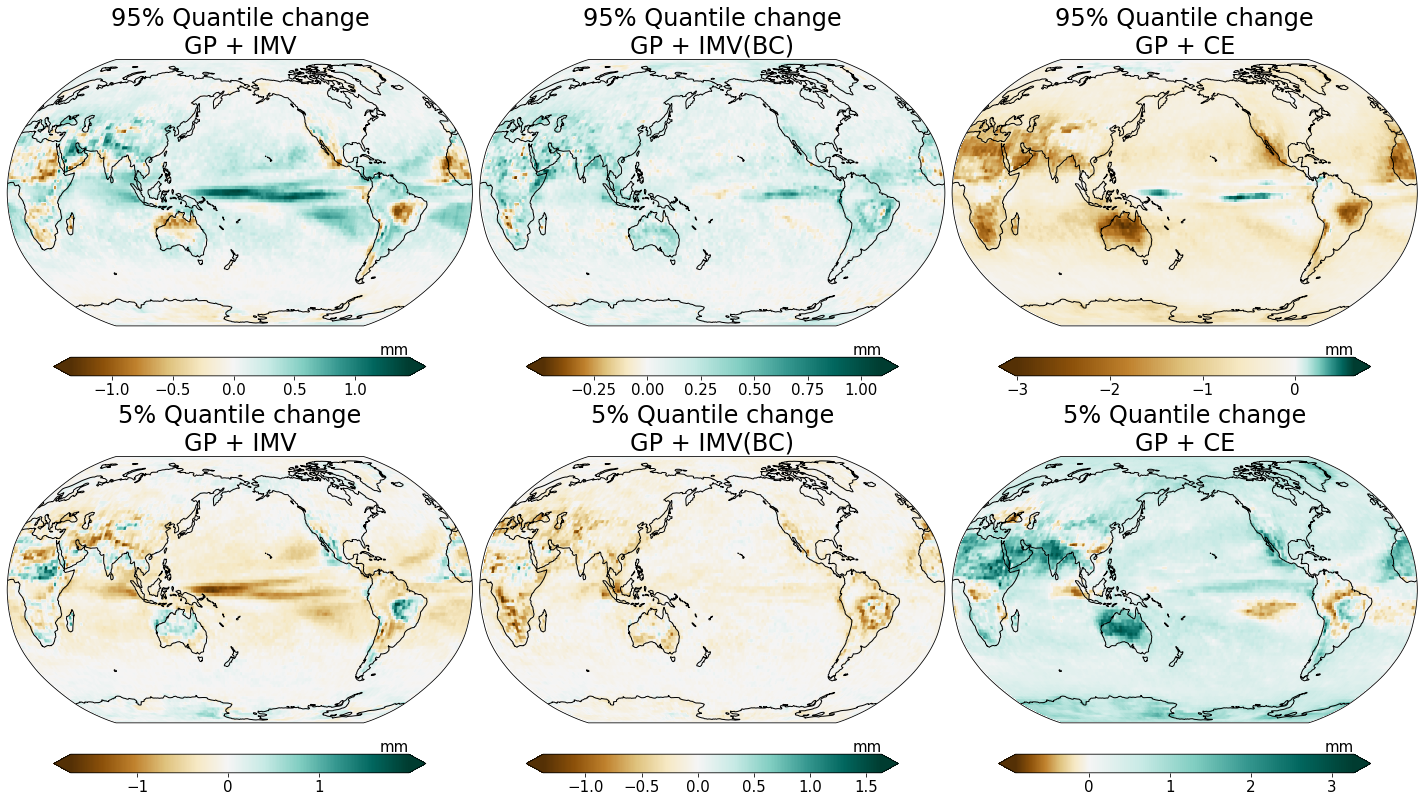} 
\caption{\textbf{Top:} Pointwise 95\% quantile difference maps for a Gaussian process model with each of the three UQ methods on monthly total precipitation (PR) data. Maps show the difference between the projected 95\% quantile for PR over the period 2024-2054 and the historical 95\% quantile for PR over the period 1960-1990 at that location. \textbf{Bottom:}  Same as the top row but using 5\% quantiles instead. Green areas indicate the 95\% or 5\% quantile has increased relative to the baseline historical period, and brown areas indicate a decrease.}
\label{fig:pr_extremes}
\end{center}
\end{figure}

Figure \ref{fig:pr_extremes} shows the quantile change plots for PR using the three projection methods. Unlike the TAS quantile change plots (Figure \ref{fig:tas_extremes}), where GP + IMV and GP + CE largely agree, GP + CE shows a markedly different projection for the tails of the PR distribution. Comparing the 95\% and 5\% quantile maps of GP + CE shows that the upper end of the PR distribution is broadly expected to decrease, while the lower end is broadly expected to increase. GP + IMV shows overall slight increases in the 95\% quantile and slight decreases in the 5\% quantile. Thus, the CE is projecting a narrowing of the PR distribution, while IMV is projecting a slight widening.
As in TAS, the conformal results match physical arguments that suggest wet areas should, generally, become wetter and dry areas should become drier \citep{ipcc2023ch4}. However, climate models struggle to simulate these changes due to their lower resolution and inability to capture processes contributing to precipitation extremes without downscaling and bias correction \citep{tapiador2019precipitation}.

\subsection{Global mean projections and uncertainty}

The numerical experiments in Section \ref{sec:experiments}, particularly Figures \ref{fig:temporal_coverage} and  \ref{fig:temporal_stability}, give us some confidence that CE can produce reliable projection uncertainty even far into the future (2070-2100). Historical validation on reanalysis data (Section \ref{sec:historical_validation}) shows that these results hold, at least in the early part of the projection period, when targeting quasi-observational data. We now use our method to project TAS and PR all the way through 2100. Again, we train on 1940-2004 data for TAS and 1940-1990 data for PR and construct CEs on 2005-2024 data for TAS and 1991- 2024 data for PR. This time, however, we will focus on the projected distribution of the global mean of each climatic variable using IMV, IMV(BC), and CE.

\begin{figure}[!t]
\begin{center}
\includegraphics[width=0.975\textwidth]{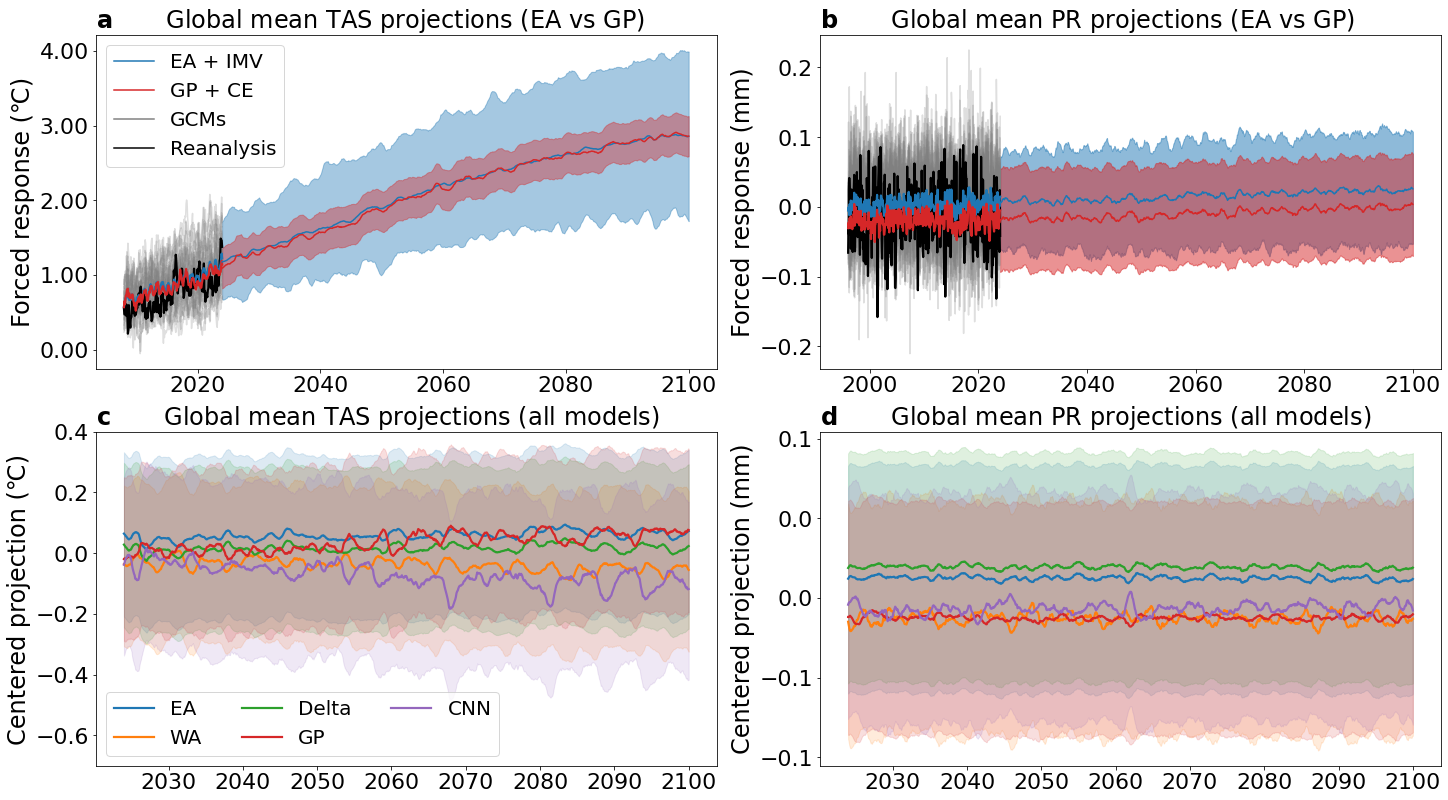} 
\caption{Panels (\textbf{a}) and (\textbf{b}) show the projected global mean monthly average surface temperature (TAS) and global mean monthly total precipitation (PR) using an ensemble average with inter-model variability based bands (EA + IMV) and a Gaussian process with conformal ensemble based bands (GP + CE). Panels (\textbf{c}) and (\textbf{d}) show the de-meaned projections, with CE based bands, for all models, excluding LM due to its outlying poor performance in numerical experiments. Projections are de-meaned by subtracting the mean projection at each time point. For all panels, the projections and uncertainty bands are smoothed using a 12-month moving average to reduce seasonal variability in the plot.}
\label{fig:reanalysis_projections}
\end{center}
\end{figure}

Figure \ref{fig:reanalysis_projections}a,b show the projected distributions of the global mean average monthly surface temperature and total monthly precipitation. The projection lines and 90\% projection bands are slightly smoothed using a 12-month moving average to remove seasonal variability. We compare the projected distributions of EA + IMV, the standard approach, against GP + CE, one of our proposed approaches. For TAS, both EA + IMV and GP + CE project almost the exact same global mean, but the 90\% projection region for GP + CE is much tighter than the 90\% projection region of EA + IMV. For PR, EA + IMV and GP + CE have nearly the same width, but GP + CE projects a slightly drier climate. These figures illustrate an important component of conformal methods. For variables with a high degree of predictability, such as TAS, a CE can result in significantly tighter uncertainty estimates than the naive IMV approach. For other variables with a lower degree of predictability, such as PR, a conformal method is not guaranteed to automatically decrease the width of the projected distribution. 

Figures \ref{fig:reanalysis_projections}c, d show the projections of all ensemble analysis methods with CE based 90\% projection region, centered across all projections at each time point, to highlight the differences between different ensemble analysis functions. For TAS, there is broad agreement between the global mean projections of each method, with the exception of the CNN, which drifts downward in the later half of the projection period. This may reflect the relatively poor robustness of CNN methods to covariate shift \citep{schneider2020improving}. Otherwise, all methods produce projections and projection regions that are nearly identical to each other. For PR, the analysis functions are split between projecting relatively wet climates (EA and Delta) or relatively dry climates (WA, GP, CNN). Figures  \ref{fig:reanalysis_projections}c, d show why it may be important to consider CE projections from multiple models when assessing future climate variability.

\section{Discussion and future work}

Climate model projections are essential tools for understanding and quantifying the risks associated with climate change, and multi-model ensembles are the primary means for quantifying projection uncertainty. Large multi-model ensembles allow us to characterize the distribution through inter-model variability (IMV), which reflects a range of plausible outcomes. We introduced a new conformal ensemble approach that allows us to condition large multi-model ensembles on quasi-observational data to improve their uncertainty quantification (UQ) capabilities. This approach works by defining an ensemble analysis function, which uses the multi-model ensemble to predict observational data and a data depth function to identify the typical sets of prediction residuals. We use this approach to construct $(1-\alpha)$-level projection sets by identifying a central region of prediction residuals (Section \ref{sec:conf_ensembles}) and adding these residuals on to the projections of the ensemble analysis function. This approach is fast and computationally efficient, requiring only a single model fit and evaluation of a depth function over a small calibration set. 

A major advantage of the conformal approach is that we do not require the data to be independent and identically distributed to rigorously guarantee the validity of the prediction sets. As long as the data are exchangeable, the bounds in Equation \ref{eqn:coverage} will hold. This makes our method well suited to applications in climate simulations and reanalysis fields, which often exhibit strong spatiotemporal correlations. 
However, spatial and temporal correlations can still impact the empirical performance of our approach. We implicitly assumed that the data exhibit strong enough spatial structure to be treated as continuous surfaces so that functional data depth methods can be applied. Multi-model analysis functions that exploit spatial structure (E.g. CNNs and GPs) are likely to show higher accuracy and, therefore, more precise prediction sets. Temporal correlations could potentially impact the effective sample size of the calibration set, which could lead to excessive over-coverage (Equation \ref{eqn:coverage}), a phenomenon that we empirical observed (Table \ref{tab:coverage}).

In Section \ref{sec:experiments}, we evaluated the out-of-sample coverage of the theoretical conformal prediction set (Table \ref{tab:coverage}) and the uncertainty quantification skill of the conformal ensemble (CE) (Table \ref{tab:uq_metrics}) across a variety of ensemble analysis functions, depth functions, and climatic variables. Our results showed that, regardless of the ensemble analysis function and climatic variable, the CE based projection regions more closely match the joint distribution of the target process than either the IMV or quantile-corrected IMV (IMV(BC)). These results were seen to hold over all time periods in the projection period (Figure \ref{fig:temporal_stability}). We also investigated the marginal UQ skill of the CE and found that in TAS and TMAX, it consistently improved over the IMV and IMV(BC), was largely similar to the UQ skill of the IMV in PR while being significantly better than the UQ skill of IMV(BC) in PR (Figure \ref{fig:spatial_stability}).

In Section \ref{sec:application}, we applied our method to reanalysis data and found, through a small historical validation experiment (Section \ref{sec:historical_validation}) that the benefits of CE seen on numerical experiments in Section  \ref{sec:experiments} likely translate to real climatic data. We then used our method to project monthly average surface temperatures (TAS) and monthly total precipitation (PR) over the next 30 years (2024-2054) and compared the projected changes in the 95\% and 5\% quantiles of each variable against historical reference data (Figures \ref{fig:tas_extremes} and \ref{fig:pr_extremes}). For TAS, we found that both the lower and upper quantiles of the distribution are projected to increase over the reference period, particularly in the Arctic, while an IMV-based approach showed a significant widening of the quantiles compared to the reference period. PR showed that CE projects a largely drier climate than the IMV, although this can depend on the chosen ensemble analysis function. Finally, figure \ref{fig:reanalysis_projections} demonstrated the precision of CE intervals depends on the target climate variable but that different ensemble analysis functions return very similar projections and interval widths for macroscopic summary statistics like the global mean.

Although our results showed that conformal ensembles have generally improved UQ skill compared to inter-model variability (even when bias-corrected), we view CE as a complementary approach to inter-model variability for climate uncertainty quantification. The IMV and a CE can be useful for assessing a multi-model ensemble's unconditional and conditional uncertainty on a given projection task. Unconditional uncertainty is intrinsically useful because it allows us to quantify the models' uncertainty. Comparing the two might be useful for investigating the predictability of different climatic variables \citep{delsole2004predictability}. CE could even be applied to single-model large initial conditional ensembles to investigate the internal variability of individual models. Focusing on a single model would allow us to isolate internal variability from other sources of variation such as inter-model variability.

Furthermore, our conformal approach is intrinsically tied to the selected ensemble analysis function, depth function, calibration set, and even the reanalysis data product used. Our results showed that the choice of ensemble analysis function may not play a huge role, but more accurate functions generally result in tighter projection sets (Table \ref{tab:coverage}). The choice of depth function does play a role in the robustness of the projection sets to violations of exchangeability, and proper depth functions (e.x. $\ell_{\infty}$-Depth or Tukey depth) should be preferred (Figure \ref{fig:temporal_coverage}) over more typical scoring rules such as the $\ell_\infty$-Norm. Our sensitivity analysis showed that the calibration set can greatly affect UQ skill and that larger sets are not necessarily better (TAS and TMAX) due to the non-stationary nature of many climatic variables. Finally, if we swap ERA5 reanalysis data for a different reanalysis data product, such as NCEP \citep{kalnay1996ncep} then our projections will likely change. Exploring the potential for using multiple observational datasets, or higher frequency (daily) data, to constrain model ensembles in this framework remains an interesting direction for future work. Otherwise, in practice, tuning the calibration set size and considering multiple reanalysis data products may be necessary to marginalize these effects.

Finally, the conformal approach can be limited because it does not attempt to model uncertainties in representing climate system physics. Because climate variables are often non-stationary and hence non-exchangeable, this could lead to poorly calibration projection sets, particularly in the future. However, these uncertainties can be incorporated through the ensemble analysis function. For example, we could, in principle, conformalize the projections of one of the Bayesian approaches, although this is currently computationally unattractive. In future work, we will consider efficient approaches for directly incorporating modeling uncertainties into general ensemble analysis functions to improve the adaptability and reliability of the conformal ensembles. We may also consider stochastic sampling schemes to generate even larger ensembles with improved representation of the theoretical projection set.

\begin{acks}[Acknowledgments]
The authors would like to thank the anonymous referees, an Associate
Editor and the Editor for their constructive comments that improved the
quality of this paper.
\end{acks}

\begin{funding}

\end{funding}

\begin{supplement}
	\stitle{Appendix}
	\sdescription{Additional tables, figures, and details are included in the Appendix.}
\end{supplement}

\begin{supplement}
\stitle{Code}
\sdescription{Contains Python code implementing our model and scripts that allow for reproducing paper results. Includes scripts for processing data, fitting all models, and creating all plots and tables. GitHub Respository: \url{https://github.com/trevor-harris/confensemble}}
\end{supplement}

\bibliographystyle{imsart-nameyear}
\bibliography{conform}

\end{document}


\begin{frontmatter}
\title{Appendix: Multi-model Ensemble Analysis with Neural Network Gaussian Processes}
\runtitle{Climate Model Integration}

\begin{aug}
\author[A]{\fnms{Trevor} \snm{Harris}\ead[label=e1,mark]{tharris@tamu.edu}},
\author[B]{\fnms{Bo} \snm{Li}\ead[label=e2]{libo@illinois.edu}}
\and
\author[C]{\fnms{Ryan} \snm{Sriver}\ead[label=e3]{rsriver@illinois.edu}}
\address[A]{Department of Statistics,
Texas A\&M University,
\printead{e1}}

\address[B]{Department of Statistics,
University of Illinois at Urbana Champaign,
\printead{e2}}

\address[C]{Department of Atmospheric Sciences,
University of Illinois at Urbana Champaign,
\printead{e3}}

\end{aug}
\end{frontmatter}

\appendix
\section{Parameter Estimation} \label{sec:prior_optimization}

We use $\theta = \{\sigma^2_w, \sigma^2_b \}$ to denote the weight and bias variances in the neural network, and $\phi = \{\theta, \sigma^2 \}$ to denote the full parameter collection including the error variance. For simplicity, and without loss of generality, we will assume $X\beta$ is zero. Because we further assume $\phi$ is shared across all spatial locations and all locations are independent, we can write the joint distribution over $Y_i(p)$ for all $i \in 1,...,n$ and $p \in s_1,...,s_d$, as
\begin{equation}\label{eqn:joint_likelihood}
   N(0, \Sigma(\theta)_{n \times n} \otimes I_d + \sigma^2 I_{nd}),
\end{equation}
where $\otimes$ denotes the Kronecker product. To maximize the likelihood, we take the negative log-likelihood, which is proportional to
\begin{equation} \label{eqn:original_loss}
	\mathcal{L}(\phi, D) = \frac{1}{2}Y^T K^{-1}(\phi)Y + \frac{1}{2} \log|K(\phi)|.
\end{equation}

For most practical problems, the dimension of the reanalysis fields, $d$, is too large to compute the true $nd \times nd$ covariance matrix $K(\phi)$, let alone invert $K(\phi)$. However, we can avoid unnecessary computations and make matrix inversion manageable by exploiting the block diagonal structure of the covariance matrix. That is, We can make evaluations of $\mathcal{L}(\phi, D)$ tractable by using the fact that
\begin{align*}
	K(\phi) &= \Sigma(\theta) \otimes I_d + \sigma^2 I_{nd} \\
	          &= (\Sigma(\theta) + \sigma^2 I_{n}) \otimes I_d,
\end{align*}
since we assumed each location has the same noise distribution  $N(0, \sigma^2)$.
Therefore, $K^{-1}(\phi) = (\Sigma(\theta) + \sigma^2 I_{n})^{-1} \otimes I_d^{-1}$ and 
\begin{align*}
	Y^T K^{-1}(\theta)Y = \sum_p  Y^T_p (\Sigma(\theta) + \sigma^2 I_{n})^{-1} Y_p,
\end{align*}
where $Y_p$ is a $1 \times n$ vector consisting of the $p'th$ location in each of the $n$ reanalysis fields. For our model integration problem, $n$ is typically small enough ($n < 1000$) to quickly invert the $n \times n$ matrix $(\Sigma(\theta) + \sigma^2 I_{n})$. Similarly, we use $K(\phi)  =(\Sigma(\theta) + \sigma^2 I_{n}) \otimes I_p$, to get
\begin{align*}
	\log|K(\phi)| &= \log|(\Sigma(\theta) + \sigma^2 I_{n}) \otimes I_p| \\
	       &= p\log|\Sigma(\theta) + \sigma^2 I_{n}| + n \log|I_p| \\
	       &= p\log|\Sigma(\theta) + \sigma^2 I_{n}|.
\end{align*}
These two manipulations result in an equivalent loss function (to equation \ref{eqn:original_loss})
\begin{equation} \label{eqn:efficient_loss}
	\mathcal{L}(\phi, D) = \sum_p  Y^T_p (\Sigma(\theta) + \sigma^2 I_{n})^{-1} Y_p + p\log|\Sigma(\theta) + \sigma^2 I_{n}|,
\end{equation}
with a significantly lower memory cost.

The loss in Equation \ref{eqn:efficient_loss} does not readily admit closed-form solutions for any parameter in $\theta$. However, for some common activation functions, such as ReLU or Tanh, $\mathcal{L}(\theta, D)$ can be differentiated with respect to $\theta$, since $\Sigma(\theta)$ is differentiable with respect to $\theta$. We can, therefore, apply gradient descent or any of its variants to find local minimizers $\hat{\theta}$. Stochastic gradient descent is also possible with mini-batching performed over the $p$ locations rather than the $n$ samples.

Optimizing the prior, rather than setting it, is crucial since, as shown in \citep{lee2017deep}, the two priors determine how well the neural network $\Phi_A(\cdot, \theta)$ can learn. The variance parameters and the activation function define a phase space, with large regions corresponding to poor predictive performance. Manually setting the parameter values is precarious since the critical region where training can occur can be small and shrinks with increasing depth. We optimize their values to locate this critical region to help ensure our model (Equation 7 in the manuscript) has low approximation and generalization error.

\section{Additional Experiments} \label{sec:additional_experiments}

\subsection{Lagged Inputs} \label{sec:lagged_inputs}

Notably, our model (Equation 7 in the manuscript) does not explicitly account for shifts in time, i.e. the influence of past or future ensemble values. However, we can accommodate this type of temporal dependence by concatenating the ensemble at time $t$ with the ensemble at time $t+1$, $t-1$, $t-2$, etc and using the concatenated ensembles as inputs to NN-GPR. Lagged ensembles provide additional information and could potentially improve predictive performance.

    \begin{figure}
    \includegraphics[width=1\textwidth]{experiments/lagged_differences.png}
    \caption{Boxplots of the difference between NNGPR with no lagged ensembles (lag0), a single backward time lag $t-1$ (lag1), and one backward and one forwards lag $t-1$ and $t+1$, respectively (lag2). In the lagged cases the contemporary ensemble at time $t$ is also included.}
    \label{fig:lagged_differences}
    \end{figure}

We conducted a small simulation study to compare NN-GPR's performance with and without lagged models for T2M prediction. We compare three settings:
    \begin{itemize}
        \item[1.] lag0 -- NN-GPR using the ensemble from time $t$ input (standard NN-GPR)
        \item[2.] lag1 -- NN-GPR using the ensembles from time $t$ and $t-1$ as input
        \item[3.] lag2 -- NN-GPR using the ensembles from time $t+1$, $t$, and $t-1$ as input
    \end{itemize}
Using each of these three models, we replicated the T2M portion of the perfect model experiments and summarize the results with MSE (Figure \ref{fig:lagged_differences}). Figure \ref{fig:lagged_differences} shows no MSE improvement by including lagged ensembles either by decade or by model being predicted. The unlagged version (lag0) may even slightly improve over lag2 as we predict farther into the future.

\subsection{Hyperparameter Sensitivity}  \label{sec:hyperparameter_sensitivity}

As mentioned in Section 3, there are several hyperparameters that govern the behavior of the NNGP, such as depth, layer type, and activation function. We conducted a sensitivity analysis to study the effect of network depth on our results. 

We repeat the T2M perfect model experiments (Section 4) with 20 different NN-GPR models by varying the depth from 1 to 20. We computed the MSE of NN-GPR's predictions under each depth setting (Figure \ref{fig:sensitivity} right side) and averaged this over the 16 perfect model runs. This allows the effect of depth on forecasting skill over long time scales (roughly 80 years). We then repeat our T2M reanalysis experiments (Section 5), again with 20 different NN-GPR models by varying the depth from 1 to 20. We computed the MSE for each depth setting (Figure \ref{fig:sensitivity} left side). This allows us to asses the effect of depth on short term forecasting skill (6 years) under realistic settings.

\begin{figure}
    \centering
    \includegraphics[width=0.95\textwidth]{experiments/sensitivity.png}
    \caption{\textbf{Left}: Reanalysis MSE by network depth. \textbf{Right}: Average (over 16 experiments) perfect model MSE by network depth.} \label{fig:sensitivity}
\end{figure}

In both cases, there is a definite trend (either positive or negative) with layer depth, but the spread between MSE from best to worst mode is not drastically large (Figure \ref{fig:sensitivity}). In the model experiments, we found a clear, though small, positive correlation between depth and MSE. Thus higher depth values resulted in higher MSEs for long term forecasting. Here the optimal setting was depth = 3. In the reanalysis experiments, found a slight negative trend, meaning higher depth values resulted in lower MSEs for short term forecasting. The differences were essentially negligible, but the lowest MSE setting was depth = 12. We also tested short term prediction with perfect model experiments (not shown) and found the same result as in the reanalysis predictions. 

We chose a depth of 10 as a compromise between short term prediction skill, long term prediction skill, and computational effort. In Section B, we compare the performance of NN-GPR using a range of depths from 1 to 20. We find that predictive skill is relatively insensitive to network depth although shallower networks perform slightly better for long term prediction and deep networks perform slightly better short term prediction.

\subsection{Comparison between NN-GPR and GPSE} \label{sec:nngp_gpse_comparison}

Because NN-GPR and GPSE are relatively close in performance in Table 1, we investigate their systematic differences more closely. We investigate the MSE difference between NN-GPR and GPSE \textit{within} each perfect model experiment. That is we want to see, for any given experiment or prediction period, if NN-GPR will have a lower MSE and higher SSIM than GPSE.

 \begin{figure}
    \includegraphics[width=1\textwidth]{experiments/gpse_sgpr_tas.png}
    \includegraphics[width=1\textwidth]{experiments/gpse_sgpr_pr.png}
    \caption{Boxplots of the difference between GPSE and NN-GPR for T2M and PR. Distribution of the differences shows a clear bias. GPSE always has higher MSE in T2M and usually higher MSE in PR. GPSE always has lower (worse) SSIM in T2M and PR.} \label{fig:within_experiment_metrics}
\end{figure}

Figure \ref{fig:within_experiment_metrics} is broken down by prediction decade (1-8), experiment number (1-16), and variable type (T2M and PR). Figure \ref{fig:within_experiment_metrics} shows MSE differences by decade (left column) and by model (right column) for temperature (top two rows) and precipitation (bottom two rows). For temperature it is clear that for almost every decade and every model, GPSE has a higher MSE and lower SSIM as evidenced by the boxplots being completely over (for MSE) or under (for SSIM) the blue 0 line. The only exception is in the first two decades where the MSE differences are comparable. For precipitation, the results are less unanimous, but show that GPSE generally has a higher MSE (by model and year) and a uniformly lower SSIM (by model and year). We conclude that, while the average metrics are close in Table 1, if we look at any given prediction task or prediction time period that NN-GPR shows a small but clear advantage.

\subsection{Miscalibration Error} \label{sec:Miscalibration Error}

Figure \ref{fig:pit} shows the miscalibration error ($L_2$ distance between PIT and a uniform density) for each method except the CNN. NN-GPR is the least calibrated for temperature prediction, and comparable with GPSE and GPEX for precipitation. Some miscalibration is likely due to NN-GPR having constant variance leading to simultaneous over and under estimation of the variance in different regions. To test this idea we created NN-GPR (cal) which uses the predictions from NN-GPR and the variance from LM. The calibration of NN-GPR (cal) is comparable to the approaches with spatially varying variance (LM, EA, and WEA). Ideally, we would incorporate spatially varying variance directly in our model but the direct likelihood approach is computationally infeasible (covariance matrix increases from $n \times n$ to $nd \times nd$ for $d$ spatial locations), so we leave this to future efforts. Miscalibration could also be due in part to the ReLU activation leading to overconfidence \cite{kristiadi2020being}, although we do not prove this here.

\begin{figure}
\centering
\includegraphics[width=0.9\textwidth]{figs/pit2.png}
\caption{Miscalibration error as measured by PIT for the NN-GPR, LM, EA, WEA, GPSE, and GPEX. We also include NN-GPR (cal) which uses the predictions from NN-GPR and the predictive variance from LM. NN-GPR shows the poorest average calibration on temperature predictions, and is comparable with GPSE and GPEX on precipitation. Calibration is measured on reanalysis field predictions from 2015-2021 (same data as Section 5).} \label{fig:pit}
\end{figure}

\bibliographystyle{imsart-nameyear}
\bibliography{nngp}